\documentstyle[a4,12pt,pstricks,pst-node,epsf,amssymb]{article}

\input{tcilatex}
 
\begin{document} 

\author{Massimo Di Pierro \\ 
{\footnotesize Theory Group - Department of Physics and Astronomy}\\ 
{\footnotesize University of Southampton - SO17 1BJ - United Kingdom}\\ 
{\footnotesize Email: {\tt mdp@hep.phys.soton.ac.uk}}} 
\title{{\tt MDP\_QCD}: Object Oriented Programming \\
for Lattice Gauge Theory  \\
{\normalsize Make lattice simulations on your home PC}} 
\date{16-10-1998} 
\maketitle 
 
\begin{abstract} 
This is a manual (built by examples) to explain the use of {\tt
MDP\_QCD}
\footnote{
The code described can be dowloaded from the web page: 
\vskip 0.3cm
{\tt
www.hep.phys.soton.ac.uk/hepwww/postgrad/M.DiPierro/software.html} 
\vskip 0.3cm
It can be freely copied and used as long the name of the author is
retained.\\
The author declines any responsibility for any improper or
unauthorized use of this software. \\
If you find any bug in the code, or you have any suggestion, please
report it to the author.}.  
It consists of an ensemble of classes and functions 
(written in GNU {\tt C++}) 
to help in writing programs for lattice QCD
in a particularly Object Oriented fashion. Some tricks are
implemented, hidden in the class definition, to optimize speed
and reduce memory usage on PCs, workstations and parallel computers with
sheared memory. 
\end{abstract}

\newpage
\tableofcontents
\newpage

\section{Introduction}

This paper describes {\tt MDP\_QCD}, a 
collection of classes and functions written in GNU\footnote{
It has been tested with GNU {\tt gpp} compiler on Solaris, Linux and
Windows NT. Modifications may be needed for a different compiler.
} {\tt C++} \cite{core}\cite{bjarne} to provide a 
framework to develop fast and efficient code for lattice simulation of $%
SU(N_c)$ gauge theories (with Wilson \cite{wilson} or Sheikoleslami-Wolhert
\cite{symanzik}\cite{sw} 
fermions). For references see
\cite{creutz}\cite{rothe}\cite{montvay}

The main characteristics of {\tt MDP\_QCD} are: 
 
\begin{itemize} 
\item  Implementation of the standard mathematical language used by 
physicists. 
 
\item  It is based on {\tt Toy\_class} and {\tt Matrix} (a class for 
matricial manipulations), both written by the author. The assignment 
operator and the copy constructor have been overloaded in a new way to 
reduce the necessity of copying when an object containing a 
pointer to dynamically allocated memory is returned by a function. 
 
\item  Four main classes: {\tt gauge\_field}, {\tt pl\_field}, {\tt
 em\_field} and {\tt fermi\_field}, all
based on the same optimization trick used in {\tt Toy\_class}. The basic 
binary operators have been overloaded for these classes. 
 
\item  An advanced random number generator, including a generator of random $%
SU(N_c)$ matrices. 
 
\item  Standard simulation algorithms: a quenched {\tt multihit()} 
Metropolis montecarlo and a {\tt mul\_invQ()} to invert the fermionic matrix.

\item  Implementation of stochastic all-to-all propagators for light
quarks.

\item  An optimized compressor for saving the gauge configurations. 
  
\item Everything works for an arbitrary number of colors, $N_c$. 
\end{itemize} 
 
The main idea on which the code has been built is the solution of the 
problem of returning objects containing pointers to dynamically allocated 
memory (realized in {\tt class Matrix},{\tt \ Toy\_class } and inherited by 
{\tt gauge\_field}, {\tt pl\_field}, {\tt em\_field} and {\tt fermi\_field}). 
 
Consider the following code: 

{\footnotesize \begin{verbatim} 
      Matrix A,B(10,10); 
      A=B; 
      A=B*A; 
\end{verbatim}} 
 
In the first assignment one wants each element of {\tt B} to be copied in the 
corresponding element of {\tt A}. 
In the second assignment it is faster if {\tt B} and {\tt A} 
are passed to the local variables of the {\tt operator*} by reference 
(i.e. without copying all the elements). 
Moreover one wants the local variable created by the {\tt operator*} to 
occupy the same memory location as the variable that is returned (this avoids 
copying and wasting space). 
To implement this idea each {\tt Matrix} object contains a {\tt FLAG} and a 
pointer to dynamically allocated memory (where the numbers are stored). 
The copy constructor and the {\tt operator=} have been overloaded in such a 
way to take care of the status of the {\tt FLAG} and eventually to copy the 
pointer to the memory, instead of copying 
the memory containing the real matrix. 
 
A physical location of memory may be pointed by different {\tt Matrix} 
objects, but this never generates confusion if the safety rules (stated 
in appendix D) are followed. An automatic system of garbage collecting deallocates 
the unused memory when there are no objects alive pointing to it. 
 
In this way, in the first assignment of the example, 
$11+800$ bytes\footnote{%
The number $11$ is the size in bytes of a {\tt Matrix} object. In is
independent from the size of the memory occupied by the ``real''
matrix.} are copied, 
while in the second assignment only $11$ bytes are copied three times (when 
passing {\tt A} and {\tt B} to the {\tt operator*()} and 
when returning the result) to 
be compared with the $800$ bytes of the matrix. The pointer of {\tt A}
 is swapped 
only when the multiplication is terminated without generating confusion 
between the input and the output of the function. This is faster than it 
would be possible in FORTRAN. In FORTRAN it would be necessary to create a 
temporary array where to store the result of the multiplication and then
copying it into {\tt A}. 
 
The {\tt FLAG} takes care of everything and the procedure works in every 
possible recursive expression.
 
One more optimization is contained in the following code: 

{\footnotesize \begin{verbatim} 
       gauge_field U; 
       fermi_field psi; 
       psi(x)=U(x,mu)*psi(x); 
\end{verbatim}} 
In this example {\tt U(x,mu)} and {\tt psi(x)} are objects of {\tt class 
Matrix} but each of them, instead of containing a pointer to new dynamically 
allocated memory,
contains pointer to the same location of memory allocated by the  
constructor of {\tt U} or {\tt psi}.
In other words: every time the programmer introduces a new object, the 
compiler knows if it is necessary to allocate new memory for it or just 
superimpose it to an existing allocated portion of memory.

All the details about these tricks are hidden from the high level 
programmer, who does not even need to know what pointers are.

\section{{\tt MDP\_Lib1.h} and its usage}

Five classes are declared in the file {\tt MDP\_Lib1.h}
\footnote{%
The classes defined in the file {\tt MDP\_Lib1.C} have already been used 
in some analysis programs written by the Southampton Theory Group \cite{me1}%
\cite{me2}\cite{me3}\cite{me4}.}:
\begin{itemize}

\item {\tt Complex}. Declared as {\tt complex<float>}. The imaginary
unit is implemented as a global constant {\tt I=Complex(0,1)}. 

\item {\tt Toy\_class}. It has no explicit application. It contains
all the tricks to minimize the memory usage and maximize speed. 
Its members are inherited by the fields defined in {\tt MDP\_QCD.h}. 
The programmer does not need to know about the structure of {\tt
Toy\_class} and for this reason its description is postponed to 
appendix C and D.

\item  {\tt Matrix}. An object belonging to this class is effectively
a complex matrix and it may have arbitrary dimensions. It is based on the
same optimization techinique described in appendix C and D.

\item {\tt Random\_generator}. This class contains public member
functions to generate random {\tt long}, {\tt float}, {\tt double}
 numbers and random
$SU(n)$ matrices. {\tt Random} is a global object belonging to this
class and it can be used to gain access to any member function.

\item {\tt JackBoot}. It enables to compute Jackkinfe and Boostrap errors
in a particularly easy and general Objcet Oriented way.
\end{itemize}

\subsection{\tt class Matrix}

A matrix object, say {\tt M}, can be declared in two different ways 
{\footnotesize \begin{verbatim} 
     Matrix M(r,c);     // r rows times c columns 
     Matrix M;          // a general matrix 
\end{verbatim}} 
 
\noindent 
Even if the size of a matrix has been declared it can be changed 
anywhere with the command

{\footnotesize \begin{verbatim} 
     M.dimension(r,c);  // r rows times c columns 
\end{verbatim}} 

\noindent 
Any matrix is automatically resized, if necessary, when a value is
assigned to it. The following lines are correct and print {\tt 8,8}

{\footnotesize \begin{verbatim} 
     Matrix M(5,7), A(8,8);
     M=A;
     printf("%i,%i\n", M.rowmax(),M.colmax());
\end{verbatim}} 

\noindent 
The memeber functions {\tt rowmax()} and {\tt colmax()} return
respectively the number of rows and columns of a {\tt Matrix}.

The element $(i,j)$ of a matrix {\tt M} can be accessed 
with the natural syntax

{\footnotesize \begin{verbatim} 
     M(i,j)
\end{verbatim}} 

\noindent 
Moreover the class contains functions
to perform standard operations:

{\footnotesize \begin{verbatim} 
     +, -, *, /, +=, -=, *=, /=, inv, det, exp, sin, cos, log, 
     transpose, hermitiam, minor, identity,... 
\end{verbatim}} 
 
\noindent 
As an example a program to compute 
 
\begin{equation} 
\left[ \exp \left( \matrix{ 2 & 3 \cr 4 & 5\mathit{i} } \right) \right]^{-1} 
\end{equation} 

\noindent 
is the following

{\footnotesize \begin{verbatim}
     // Program saved in FILE: p01.C
     #include "MDP_Lib1.h" 
     int main() { 
        Matrix a(2,2); 
        a(0,0)=2; a(0,1)=3; 
        a(1,0)=4; a(1,1)=5*I; 
        print(inv(exp(a))); 
        return 0; 
     }; 
\end{verbatim}} 
 
\noindent
It is straightforward to add new functions copying the prototype
\footnote{
A list of general safety rules to be applied 
when using the class {\tt Matrix} can
be found in Appendix D.
}

{\footnotesize \begin{verbatim} 
     Matrix f(Matrix a, Matrix b, ...) { 
        Matrix M; 
        // body, 
        prepare(M); 
        return M; 
     }; 
\end{verbatim}} 

\noindent
One more example of how to use the class {\tt Matrix} follows

{\footnotesize \begin{verbatim} 
      // Program saved in FILE: p02.C
      #include "MDP_Lib1.h" 
      Matrix cube(Matrix X) { 
         Matrix Y; 
         Y=X*X*X; 
         prepare(Y); 
         return Y; 
      }; 
      int main() { 
         Matrix A,B; 
         A=Random.SU(3); 
         B=cube(A)*exp(A)+inv(A); 
         print(A); 
         print(B); 
         return 0; 
      }; 
\end{verbatim}} 

\noindent 
This code prints on the screen a random SU(3) matrix $A$ and
$B=A^3e^A+A$.\\
Some example statements are listed in tab.~\ref{figmatrix1}. Note the
command

{\footnotesize \begin{verbatim} 
       A=mul_left(B,C);
\end{verbatim}}

\noindent
that is equivalent but faster than

{\footnotesize \begin{verbatim} 
       A=C*transpose(B);
\end{verbatim}}

\noindent
It will be useful to multiply a fermionic field (seen as a set of
$color \times spin$ matries) by a spin structure.

\begin{table}
\begin{center}
\begin{tabular}{|l|l|} 
\hline 
Example & {\tt C++ with MDP\_Lib1.h} \\ \hline 
$A\in M_{r\times c}({\bf C})$ & {\tt A.dimension(r,c)} \\  
$A_{ij}$ & {\tt A(i,j)} \\ 
$A=B+C-D$ & {\tt A=B+C-D} \\
$A^{(ij)}=B^{(ik)}C^{(kj)}$ & {\tt A=B*C} \\ 
$A^{(ij)}=B^{(jk)}C^{(ik)}$ & {\tt A=mul\_left(B,C)} \\ 
$A=aB+C$ & {\tt A=a*B+C} \\ 
$A=a{\bf 1}+B-b{\bf 1}$ & {\tt A=a+B-b} \\ 
$A=B^TC^{-1}$ & {\tt A=transpose(B)*inv(C)} \\ 
$A=B^{\dagger }\exp (iC)$ & {\tt A=hermitian(B)*exp(I*C)} \\
$A=\cos (B)+i\sin (B)*C$ & {\tt A=cos(B)+I*sin(B)*C} \\
$a=\func{real}(tr(B^{-1}C))$ & {\tt a=real(trace(inv(B)*C))} \\
$a={\det (B)*\det (B}^{-1})$ & {\tt a=det(B)*det(inv(B))} \\ \hline 
\end{tabular} 
\caption{Examples of typical instructions acting on {\tt Matrix}
objects.  {\tt A,B,C,D} are assumed to be declared 
as {\tt Matrix}; {\tt r,c} as {\tt int}; {\tt a,b} may be any kind of
number.} 
\label{figmatrix1}
\end{center}
\end{table}

\subsection{\tt class Random\_generator}

It is possible to declare different objects belonging to this class
but this should be done only if one needs many independent random
generators (in the sense that they relay on different seeds). 
For all the applications considered from now on, only one random
generator will be needed. To this scope a global object called {\tt Random}
is automatically declared when {\tt MDP\_Lib1.h} is included. 
Through this global object it is possible to access any member
function of the class. 
 
\begin{itemize} 

\item {\tt void Random.read\_seed(char filename[]);} It reads the seed
from the file {\tt filename} and initializes the random generator. If
no filename is provided the default filename is {\tt
RandomBuffer.seed}. If the program does not find the file with the
seed it automatically creates it with some default values in it.

\item  {\tt void Random.write\_seed(char filename[]);} It writes the
last computed seed into the file {\tt filename}. The default file name
is {\tt RandomBuffer.seed}.

\item  {\tt long Random.Long(long n);} It returns a random {\tt long} number
between {\tt 0} and {\tt n-1}.
 
\item  {\tt float Random.Float();} It returns a random {\tt float}
 number between {\tt 0} and {\tt 1}. 
 
\item  {\tt double Random.Double();} It returns a random {\tt double} 
number between {\tt 0} and {\tt 1}. 
 
\item  {\tt float Random.Gauss();} It returns a random {\tt float} 
number $x$ generated with a Gaussian probability $P(x)=\exp (-x^2/2).$ 
 
\item  {\tt float Random.Floatp(float (*P)(float, void*), void* a);} 
It returns a random 
number $x$ generated with a Gaussian probability {\tt P(x,a)}, where
{\tt a} is any set of parameters pointed by {\tt a} 
(the second argument passed to {\tt Floatp}).
 
\item  {\tt Matrix Random.SU(int n);} it returns a random {\tt Matrix}
in the group $SU(n)$ or $U(1)$ if the argument is $n=1$. 
\end{itemize} 
 
The algorithm for SU(2) is based on the observation that each rotation
in the 3D space (in the group $O(3)$) can be represented by a direction
$\widehat{ 
a}\in S^2$ and a rotation angle around that direction, $\alpha \in [0,\pi 
)$. Therefore first a random element of $O(3)$ is generated,
second, it is mapped to $SU(2)$.  The map between $O(3)$ and $SU(2)$ is realized by  
\begin{equation}
\{\widehat{a},\alpha \}\rightarrow \exp (i\alpha \widehat{a}\cdot \sigma 
)=\cos (\alpha )+i\widehat{a}\cdot \sigma \sin (\alpha )  
\end{equation}
where ($\sigma ^1,\sigma ^2,\sigma ^3$) is a vector of Pauli matrices. 
 
The algorithm for $SU(n>2)$ is the Cabibbo-Marinari~\cite{cm} iteration based 
on $SU(2)$. 
 
\subsection{\tt class JackBoot}

Suppose one has {\tt n} gauge configurations and, on each
configuration one has measured {\tt m} {\tt float} quantities 
{\tt a[0]}, {\tt a[1]},...,{\tt a[m-1]}. 
Then one wants to compute the average, over the gauge configurations,
of a function {\tt F(float *a)} with its Jackknife and Bootstrap
errors \cite{errors}.
A simple way to do it is to define a {\tt JackBoot} object, let's call it 
{\tt jb}. It is a sort of container for the data. After it has been
filled it can be asked to return the mean of {\tt F()} and its errors.
Here is an example of how it works:

{\footnotesize \begin{verbatim}
      // object declaration:
      JackBoot<float, m, n> jb;
      // assigning the function:
      jb.f=F; // note that f is a member variable
              // while F is the user defined function
      // filling the object:
      for(jb.elem=0; jb.elem<n; jb.elem++) {
         // somehow measure a[0],...a[m-1] on the gauge configs
         jb(0)=a[0];
         jb(1)=a[1];
         ...
         jb(m-1)=a[m-1];
      };
      // printing the results:
      printf("Result          = %f\n", jb.mean());
      printf("Jackknife error = %f\n", jb.j_err());
      printf("Bootstrap error = %f\n", jb.b_err(100));
\end{verbatim}}

Note that
\begin{itemize}
\item The class {\tt JackBoot<...> jb} is defined with a template where
the first argument between angular brakets 
is the basic variable type (in the example it is {\tt float}), the
second is the size of the array to be passed to {\tt f()}, the third
is the number of confugurations to be processed.

\item {\tt jb.f} is the pointer to the function used in the
analysis.

\item {\tt jb.elem} is an integer that must be used as a
counter on the configurations.

\item {\tt jb.operator(int n)} is used to access data in the
object (a configuration number in {\tt jb.elem} must be specified).

\item {\tt jb.mean()} returns the mean.

\item {\tt jb.j\_err()} returns the Jackknife error.

\item {\tt jb.b\_err()} returns the Bootstrap error.
It takes as argument the number of
Bootstrap samples. The default value is 200. 

\end{itemize}
It is important to stress that the class {\tt JackBoot} can contain
objects of any class {\tt T}, but the functions {\tt b\_err} is not 
properly defined unless
the operator {\tt T::operator>(class T)} is declared. In other
words one cannot perform a Boostrap of functions of matrices unless one
defines some kind of ordering relation between matrices.
I suggest to play safe and perform only Boostraps of {\tt float} numbers:
{\tt  JackBoot<float,n,m> }

It is possible to declare arrays of {\tt JackBoot} objects, but it is
rarely necessary. It is simpler to declare different functions and
repeat the analysis using the same {\tt JackBoot} object assigning
the pointer {\tt JackBoot::f} to each of the functions at the time. \\
As another example consider the following program. It generates an array of 100
$SU(6)$ matrices. For each matrix it computes trace and determinant, and 
returns the average of the ratio between the real part of the trace and
the real part of the determinant (with its Jackknike and Bootstrap
errors).

{\footnotesize \begin{verbatim}
      // Program saved in FILE: p03.C
      #include "MDP_Lib1.h"
      #define ng 100
      float f(float *x) { return x[0]/x[1]; };
      int main() {
         Matrix A;
         JackBoot<float,2,ng> jb;
         for(jb.elem=0; jb.elem<ng; jb.elem++) {
            A=Random.SU(6);
            jb(0)=real(trace(A));
            jb(1)=real(det(A));
         };
         jb.f=f;
         printf("Result          = %f\n", jb.mean());
         printf("Jackknife error = %f\n", jb.j_err());
         printf("Bootstrap error = %f\n", jb.b_err(100));
         return 0;
      };
\end{verbatim}}

\noindent
The output is 

{\footnotesize \begin{verbatim}
      Result          = 0.036311
      Jackknife error = 0.087398
      Bootstrap error = 0.085979
\end{verbatim}}

\section{{\tt MDP\_QCD.h} and its usage}

\subsection{Basic Syntax}

The most general {\tt C++} program using {\tt MDP\_QCD} 
must have the structure

{\footnotesize \begin{verbatim}
      #include "MDP_QCD.h"       

      // declaration of functions

      int main() {
         start();

         // initialize beta=...
         //            kappa=...
         //            c_SW=...
         // main body

         stop();
         return 0;
      };
\end{verbatim}}

The file {\tt MDP\_QCD.h} includes {\tt MDP\_Lib1.h} and {\tt
MDP\_Settings.h}. Any other header needed is included by {\tt
MDP\_Lib1.h}.

The file {\tt MDP\_Settings.h} contains the declaration of those global
constants which can be modified by the user: size of the lattice and
the number of colors. A typical settings file is

{\footnotesize \begin{verbatim}
      #define Nx0    6  // temporal sites
      #define Nx1    4  // spatial sites in the x direction
      #define Nx2    4  // spatial sites in the y direction
      #define Nx3    4  // spatial sites in the z direction
      #define Nc     3  // number of colors
      #define Nfermi 10 // used by light_propagator
\end{verbatim}}

\noindent
Note that it is sufficient to change the value of {\tt Nc } and
recompile the code to change the gauge group. No other line has to be
modified.

The instructions {\tt start();} and {\tt stop();} are compulsory and
do the job of reading/writing the seed and initializating all the
global variables and matrices. Remember that if the file containing the
seed does not exist it is automatically created.

All the global matrices are listed in Appendix A together with the 
conventions used by {\tt MDP\_QCD}. Any other built-in global variable 
is listed in (\ref{whattable}).

The basic objects in which lattice QCD is usually formulated 
are 
\begin{itemize}
\item Matrices ({\tt class Matrix})
\item Links ({\tt class gauge\_field})
\item Plaquettes ({\tt class pl\_field})
\item Chromo-electro-magnetic tensor ({\tt class em\_field})
\item Fermionic field ({\tt class fermi\_field})
\item Fermionic light propagator ({\tt class light\_propagator})
\end{itemize}

To explain the power of this language, here is an example of a
program to create one gauge configuration and test if the links
are in $SU(N_c)$ (where $N_c$ is declared in {\tt MDP\_Settings.h})

{\footnotesize \begin{verbatim}
      // Program saved in FILE: p04.C
      #include "MDP_QCD.h"       
      int main() {
         start();
         int mu, counter=0; 	
         site x;
         Matrix A;
         gauge_field U;
         U=hot();
         beta=5.7;
         multihit(U);
         for(x=0; x<Nvol; x++)
            for(mu=0; mu<4; mu++) {
               A=U(x,mu)*hermitian(U(x,mu))-1;
               if(real(trace(A*hermitian(A)))>0.0001)
                   counter++;
            };
         if (counter==0) printf("All links are in SU(N)\n");
         stop();
         return 0;
      };
\end{verbatim}}

It produces the following output:

{\footnotesize \begin{verbatim}
      =========================================================
      PROGRAM MDP_QCD. GENERAL INFORMATIONS:
      =========================================================
      Created by Massimo Di Pierro (@ 1998)
      Version n.     = 8-9-1998
      Compiling date = Oct 16 1998
      Compiling time = 20:57:58
      Num. of colors = 3
      Lattice size   = 6(t) x 4(x) x 4(y) x 4(z) = 384
      =========================================================
      PROGRAM STARTING:
      =========================================================
      Setting indices
      Defining the matrices
      Creating SU(N) generators
      Readind Random.seed
      Creating initial hot configuration...
      Multihit step n.0, beta=5.700000...
      ... Efficiency: 83.40%
      All links are in SU(N)
      Writing Random.seed
      =========================================================
      PROGRAM END.
      =========================================================
\end{verbatim}}

In the rest of this manual the follwing
declarations will be assumed

{\footnotesize \begin{verbatim}
      int              i,j,k; \\ color indices
      int              a,b,c; \\ spinor indices
      int              mu,nu; \\ lattice directions
      site             x,y,z; \\ lattice sites
      Matrix           A,B,C; \\ generic matrices
      gauge_field      U;
      pl_field         P;
      em_field         G;
      fermi_field      chi, psi, phi;
      light_propagator S;
\end{verbatim}}

\begin{table}
\begin{center}
\begin{tabular}{|l|l|l|} 
\hline 
Example & {\tt C++ with MDP\_QCD} & returned type \\ \hline 
$y=x+\mu $ & {\tt y=up[x][mu]} & {\tt site} \\ 
$y=x-\mu $ & {\tt y=dw[x][nu]} & {\tt site} \\ 
$y=x+10\widehat{\mu }$ & {\tt y=move(x,mu,10)} & {\tt site} \\
$y=x+3\widehat{\mu }+5\widehat{\nu }$ & {\tt y=move(x,mu,3,nu,5)} & {\tt site%
} \\ 
$a=\left| x-y\right| $ & {\tt a=dist(x,y)} & {\tt float} \\ 
$x=(x_0,x_1,x_2,x_3)$ & {\tt x=position(x0,x1,x2,x3)} & {\tt site} \\
$U_\mu (x)$ & {\tt U(x,mu)} & {\tt Matrix } \\ 
$U_\mu ^{ij}(x)$ & {\tt U(x,mu,i,j)} & {\tt Complex\&} \\
$\psi (x)$ & {\tt psi(x)} & {\tt Matrix } \\ 
$\psi _\alpha ^i(x)$ & {\tt psi(x,i,alpha)} & {\tt Complex\&} \\
$U_\mu ^{ij}(x)=U_\nu ^{ik}(y)U_\rho ^{kj}(z)$ & {\tt U(x,mu)=U(y,nu)*U(z,ro)%
} & {\tt void} \\ 
$U_\nu (x)U_\mu (x+\widehat{\nu })U_\nu ^H(x+\widehat{\mu })$ & {\tt %
staple\_up(U,mu,nu)} & {\tt Matrix } \\ 
$U_{-\nu }(x)U_\mu (x-\widehat{\nu })U_\nu (x+\widehat{\mu }-\widehat{\nu })$ 
& {\tt staple\_dw(U,mu,nu)} & {\tt Matrix } \\
$U_{\mu \nu}^\Box (x)$ & {\tt plaquette(U,x,mu,nu)} & {\tt Matrix } 
 \\
$\phi (x)=U_\mu (y)\psi (z)$ & {\tt phi(x)=U(y,mu)*psi(z)} & {\tt void} \\ 
$(1+\gamma ^\mu )\psi (z)$ & {\tt mul\_left(1+Gamma[mu],psi(z))} & {\tt %
Matrix } \\ 
$\sum_{x,i,\alpha }\psi _\alpha ^{i}(x)\phi _\alpha ^{i*}(x)$ & {\tt psi*phi%
} & {\tt Complex} \\
$\psi =Q[U]\chi $ & {\tt psi=mul\_Q(U,G,chi)} & {\tt void}  \\
$\psi =Q^{-1}[U]\chi $ & {\tt psi=mul\_invQ(U,G,chi)} & {\tt void} \\
$\psi =Q^{\dagger }[U]\chi $ & {\tt psi=mul\_Q(U,G,chi,,DAGGER)} &
{\tt void}  \\
$\psi =(Q^{\dagger }[U])^{-1}\chi $ & {\tt
psi=mul\_invQ(U,G,chi,,DAGGER)} & {\tt void} 
\\ \hline 
\end{tabular}
\end{center}
\caption{Example lines in {\tt C++} with {\tt MDP\_QCD}. 
Note that $Q[...]$ is the fermionic matrix derived from 
the Sheikoleslami-Wolhert action. Its explicit expression is given in
eq.~(\ref{sw}).}
\label{examples2}
\end{table}

The basic unary and binary operators have been overloaded. 
The use and power of the basic classes can be better explained with 
a the few example lines of tab.~(\ref{examples2}).

{\bf To handle properly objects belonging to any of the classes {\tt
gauge\_field}, {\tt pl\_field}, {\tt em\_field} and {\tt fermi\_field}
the safety rules described in Appendix D must be followed.}

\subsection{Creating gauge configurations}

Any standard lattice simulation begins with the creation of an ensemble of
gauge configurations $\left\{ U^i\right\} $. It is created through a Markov
process\cite{rothe}, i.e. each configuration $U^i$ is generated from
the preceding one,  $U^{i-1}$, using any stochastic algorithm  
\begin{equation}
U^i=F(U^{i-1})  \label{step1}
\end{equation}
satisfying the relation
\begin{equation}
e^{{\mathcal{S}}[U^{i-1}]}P(U^{i-1} \rightarrow U^i)=e^{-{\mathcal{S}}[U^i]}
P(U^i \rightarrow U^{i-1})
\end{equation}
where $P(U \rightarrow U')$ is the probability of generating the
configurations $U'$ from the configuration $U$. 
Here ${\mathcal{S}}[U]$ is the lattice Euclidean action evaluated on the gauge
configuration $U$. 
In {\tt MDP\_QCD} the standard quenched action is implemented 
\begin{equation}
{\mathcal{S}}[U]=\sum_x \frac \beta {N_c}\sum_{\mu ,\nu }\func{Re}\left[ \text{
tr}\left( 1-U_\mu(x)U_\nu(x+\mu)U_\mu^{\dagger}(x+\nu)
U_\nu^{\dagger}(x)
\right) \right]  \\
\label{quenched}
\end{equation}
where $\beta=1/g^2(a)$ is the parameter that fixes the lattice scale.
The action in eq.~(\ref{quenched}) is a discretized version of 
the pure Yang-Mills action, i.e. fermion loops are neglected.

The initial configuration $U^0$ can be chosen to be ``cold'',
i.e. when all its links are the identity, or ``hot'', when each link
is a random $SU(N)$ matrix.

Since it is very important to reduce the correlation between $U^{i-1}$ and $%
U^i$, it is necessary to iterate eq.(\ref{step1}) many times
\begin{equation}
U^i=F(F(...F(U^{i-1})...))
\label{stepn}
\end{equation}

Any quantum gauge observable $O[U]$ can be measured as an 
an average over the ensemble of gauge configurations
\begin{equation} 
\left\langle O[U]\right\rangle = \int [dU]O[U]P[U] \simeq
\frac 1{N_U}\sum_i O[U^i]
\end{equation} 
 
In \texttt{MDP\_QCD} a gauge configuration $U$ is an object of the class 
\texttt{gauge\_field}. It is declared with the command

{\footnotesize \begin{verbatim}
      gauge_field U;
\end{verbatim}}

\noindent
An initial hot configuration is created with the command
{\footnotesize \begin{verbatim}
      U=hot();
\end{verbatim}}

\noindent 
and an initial cold configuration is created with the command

{\footnotesize \begin{verbatim}
      U=cold();
\end{verbatim}}

The operation described in eq.(\ref{step1}) and eq.(\ref{stepn}) is
performed using the multi-hit algorithm that is described in the
section on algorithms. The command to perfrom it is

{\footnotesize \begin{verbatim}
      // beta=...
      multihit(U,num_hits,num_iter);
\end{verbatim}}

\noindent
{\tt num\_hits} is the number of hits for each multihit step. {\tt
num\_iter} is the number of multihit iterations. A value for the
gloabal variable {\tt beta} must be specified. 
The new configuration is stored in the same object $U$ which contains the
configuration which is passed as argument. This saves memory. In
principle it is possible to define arrays of configurations and define many
different gauge configurations at the same time

{\footnotesize \begin{verbatim}
      gauge_field myconf[100][3];
      gauge_field U, U1, U2, U3;
\end{verbatim}}

\noindent
and copy one into the other as if they were ordinary variables

{\footnotesize \begin{verbatim}
      myconf[37][2]=U;
      U2=U;
      U=myconf[50][0];
\end{verbatim}}

\noindent
To avoid ambiguities the assignment operator of any kind of field returns
void.

Links $U_\mu(x)$ can be accessed as matrices {\tt U(x,mu)} 
or as complex numbers $U^{ij}_\mu(x)$, {\tt U(x,mu,i,j)}. 
It is also possible to access a link as a
matrix and extract its $(i,j)$ element {\tt U(x,mu)(i,j)} but this is
slower than the direct access {\tt U(x,mu,i,j)}.

Here is a program that generates and saves an ensemble of 100 gauge
configurations, at $\beta =6,$ starting from hot, performing 20 multi-hit
steps at the time

{\footnotesize \begin{verbatim}
      // Program saved in FILE: p05.C
      #include "MDP_QCD.h"
      int main() {
         start();
         int g;
         gauge_field U;
         U=hot();
         beta=6.0;  // beta is declared in MDP_QCD.h
         for(g=0; g<100; g++) {
            multihit(U,5,20);
            write(U, "gauge_config_beta_6.0", g);
         };
         stop();
         return 0;
       };
\end{verbatim}}

\subsection{First measurements}

It only makes sense to measure gauge invariant quantities, i.e. closed loops
of links. The minimum closed loop is the plaquette ({\tt pl})

\begin{equation}
P_{\mu \nu}(x)=U_{\mu}(x)U_{\nu}(x+\widehat{\mu })U_{-\mu}(x+\widehat{\mu }+%
\widehat{\nu })U_{-\nu}(x+\widehat{\nu })
\end{equation}

\noindent
which is a rank two tensor field. It is related to the
chromo-electro-magnetic ({\tt em}) tensor field through the relation
\begin{eqnarray}
G_{\mu \nu}(x)&=&\frac18 \left[
P_{\mu \nu }(x)+P_{\mu \nu}(x-\widehat{\mu})+
P_{\mu \nu}(x-\widehat{\nu})+P_{\mu
\nu}(x-\widehat{\mu}-\widehat{\nu}) \right] \\
&-& \frac18 \left[ P_{\nu \mu }(x)+P_{\nu \mu}(x-\widehat{\mu})+
P_{\nu \mu}(x-\widehat{\nu})+P_{\nu\mu}(x-\widehat{\mu}-\widehat{\nu})
\right]
\end{eqnarray}

\noindent
In {\tt MDP\_QCD} there is a class for the plaquettes field
{\footnotesize \begin{verbatim}
       pl_field P;
\end{verbatim}}

\noindent
and a class for the chromo-electro-magnetic field

{\footnotesize \begin{verbatim}
       em_field G;
\end{verbatim}}

\noindent
For an arbitrary gauge configuration $U$ these field are computed 
with the commands

{\footnotesize \begin{verbatim}
       P=compute_plaquettes(U);
       G=compute_em_tensor(P);
\end{verbatim}}

\noindent
An interesting measurement to be performed on each gauge configuration is the
real part of the trace of the average plaquette:
{\footnotesize \begin{verbatim}
       float av_pl;
       av_pl=average(P);
\end{verbatim}}

What follows is a program that reads the gauge configurations created by
{\tt p05.C}, computes the average plaquette and prints the result for each
gauge configuration.

\begin{figure}
\epsfxsize=10cm
\epsfysize=8cm
\hfil \epsfbox{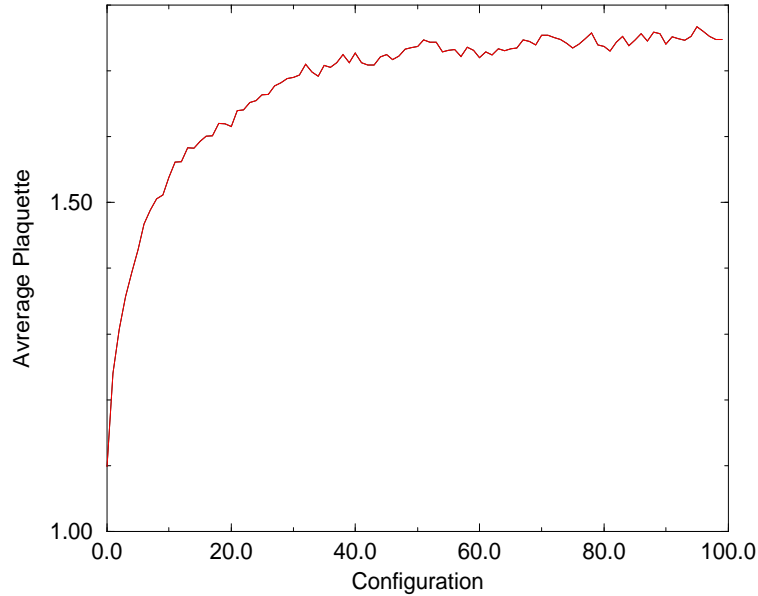} \hfil
\caption{The graph shows the real part of the trace of the average
plaquette computed on 100 gauge configurations. The plateau indicates
the termalized configurations.}
\label{avpl}
\end{figure}

{\footnotesize \begin{verbatim}
      // Program saved in FILE: p06.C
      #include "MDP_QCD.h"
      int main() {
        start();
        int g;
        gauge_field U;
        for(g=0; g<100; g++) {
          read(U, "/home/lattice/mdp/gauge_config_beta_6.0", g);
          printf("%i %f\n",g, average(compute_plaquettes(U)));
        };
        stop();
        return 0;
      };
\end{verbatim}}

\noindent
The output of the program has been plotted and is shown in
fig.~(\ref{avpl}). 

It shows that the average plaquette 
approaches an asymptotic value (which mainly depends on the value of $%
\beta $) and then  fluctuates around it. This is the region where the field is
said to be thermalized, i.e. when the statistical fluctuations compensate the
classical tendency to minimise the free energy. This is the region that
corresponds to a physical quantum vacuum. Configurations that are not
thermalized should be discarded.
As is evident from the figure, in our case the first 60 saved 
configurations are not termalized. We will refer to the remaining 40 as
``effective configurations''.

A {\tt pl\_field P} can be accessed as $N_c \times N_c$ matrix {\tt
P(x,mu,nu)} or as complex number {\tt P(x,mu,nu,i,j)}. Analogously an
{\tt em\_field G} can be accessed as an $N_c \times N_c$ matrix {\tt
G(x,mu,nu)} or as complex number {\tt G(x,mu,nu,i,j)}. {\bf In both cases
{\tt mu} must be less than {\tt nu}. The other components can be
obtained by antisymmetrizing.}

For a general review on the typical quantities that can be computed on
lattice see ref.~\cite{lists}.

As an example the next program computes a Polyakov loop on our 
effective 40 gauge configurations and prints out the result for each
gauge configuration.

{\footnotesize \begin{verbatim}
      // Program saved in FILE: p07.C
      #include "MDP_QCD.h"
      int main() {
        start();
        int g,x0;
        site x;
        gauge_field U;
        Matrix Polyakov(Nc,Nc);
        for(g=60; g<100; g++) {
          read(U, "gauge_config_beta_6.0", g);
          Polyakov=1;
          x=0;
          for(x0=0; x0<Nx0; x0++) {
            Polyakov=Polyakov*U(x,0);
            x=up[x][0];
          };
          print(Polyakov);
        };
        stop();
        return 0;
      };
\end{verbatim}}

\subsection{Introducing fermions}

Another set of observables that can be measured on the lattice are the
two and three point correlation functions between currents (and their
Fourier transform)

\begin{eqnarray}
C_2(t_x)     &=&\int d^3{\bf x} \langle 0 | J(0) J^\dagger(x) 
| 0 \rangle \label{2cp} \\
C^Q_3(t_x,t_y) &=&\int d^3{\bf x} d^3{\bf y}
\langle 0 | J(-y) Q(0) J^\dagger(x)| 0 \rangle
\label{3cp}
\end{eqnarray}

Since the lattice metric is Euclidean, the asymptotic behaviour of the
spatial Fourier transform of the two point correlation function is given
by
\begin{equation}
	C_2(t_x) \sim Z^2 e^{-m_J t_x}
\label{fit}
\end{equation}
where $m_J$ is the mass of the lightest state $| 1_J \rangle$ 
created by the current $J^\dagger$ and
\begin{equation}
	Z=\left| \langle 1_J \right| J^\dagger(0) \left| 0 \rangle  \right|
\end{equation}
Therefore from the measurement of $C_2(t_x)$ and its fit to
(\ref{fit}), it is possible to extract masses of particles.
In the same fashion from the asymptotic behaviour of the ratio between
the three and two two point correlation functions it is possible to
extract matrix elements\cite{chris}
\begin{equation}
\frac{C^Q_3(t_x,t_y)}{C_2(t_x) C_2(t_y)} \sim \frac1{Z^2} 
\frac{\langle 1_J | Q | 1_J \rangle }{2m_J}
\end{equation}

The most general current $J(x)$ can be expressed in terms of fundamental 
fermionic fields $q_\alpha ^i (x)$ (the quark fields). In 
expressions like eq.(\ref{2cp}) and (\ref{3cp}) these field are Wick 
contracted
\begin{equation}
	\langle 0 |\{ q_\alpha ^i(x), \overline{q}_\beta ^j(y) \} |0 \rangle
	=S_{\alpha \beta} ^{ij}(x,y)
\label{light}
\end{equation}
Despite the fact that fermions are neglected when gauge configurations
are created, they are reintroduced at a later stage as particles
propagating in the gluonic background field.
Therefore the two and three point correlation functions can 
be written as appropriate traces of propagators, $S_{\alpha \beta}
^{ij}(x,y)[U]$, in the backgroud gluonic field $U$.

In {\tt MDP\_QCD} fermionic fields belong to the class {\tt
fermi\_field}. 
In the following line the fields $\chi$, $\psi$ and $\phi$ are declared

{\footnotesize \begin{verbatim}
      fermi_field chi, psi, phi;	
\end{verbatim}}

\noindent
On each gauge configuration $U$, the fermion propagator $S$ is
computed by inverting the fermionic matrix 
\begin{equation}
S[U]=(Q[U])^{-1}
\label{inversion}
\end{equation}
This is the most time expensive part of any lattice calculations.
The fermionic matrix implemented comes from the
standard Sheikoleslami-Wolhert action\footnote{
The Feynman rules derived for this action can be found in
\cite{heatlie}. Note that in this paper the $\sigma_{\mu\nu}$ matrices
are different, for an {\it i} factor, from those implemented and
listed in appendix A.} \cite{luscher}
\begin{eqnarray} 
Q_{x,y}[U]&=&\delta_{x,y}-\kappa \sum_{\mu =0..3} \left[
(1-\gamma _\mu )U_\mu (x) \delta_{x,y-\mu} +
(1+\gamma _\mu )U_\mu^\dagger (x-\mu) \delta_{x,y+\mu} \right] 
\nonumber \\
&-&i \kappa c_{SW} \sum_{\mu < \nu }\sigma _{\mu \nu }G_{\mu \nu }(x)
\delta_{x,y}
\label{sw}
\end{eqnarray} 

\noindent
where $\kappa$ and $c_{SW}$ are input parameters. 
$\kappa$ is in one to one correspondence with the fermion mass
\begin{equation}
m=\frac1{2a}\left(\frac1{\kappa}-\frac1{\kappa_{crit}}\right)
\end{equation}
and $\kappa_{crit}$ is a parameter depending on $\beta$. The chiral
limit corresponds to the limit $\kappa \rightarrow \kappa_{crit}$, when
the quark becames massless. In practice any
inversion algorithm for eq.(\ref{inversion}) converges slower and
slower as the chiral limit is approached and this can never be reached. 
$c_{SW}$ also depends on $\beta$ and its purpose is to
cancel order $a$ effects in the propagator (as well as in physical
observables). $c_{SW}$ can be computed perturbatively 
(it is 1 at three level), can be extimated from the average plaquette
(equivalent to resumming ``tadpole'' contributions) or 
can be obtained by a fine tuning procedure (called ``non-perturbative''
improvement).

$\kappa$ and $c_{SW}$ are declared as global variables in {\tt MDP\_QCD}.
The multiplication of a {\tt fermi\_field} 
by the fermionic matrix is performed by the function {\tt mul\_Q()}
\footnote{
Note that before calling {\tt mul\_Q()} it is
necessary to assign a value to the global variables {\tt kappa} and {\tt 
c\_SW}}. 

{\footnotesize \begin{verbatim}
       kappa=0.135;
       c_SW=0;
       G=compute_em_tensor(compute_plaquettes(U));
       psi=mul_Q(U, G, chi);
\end{verbatim}}

\noindent
Here the second line reads
\begin{equation}
       \psi=Q[U]\chi
\end{equation}
Note that the chromo-electro-magnetic field $G_{\mu\nu}$ computed
on the gauge configuration $U$ must be passed to
the function {\tt mul\_Q}. 
This is done to speed up the computation of the so called ``clover''
term (the part of the action proportional to $c_{SW}$).

The inverse operation is performed by the function

{\footnotesize \begin{verbatim}
       chi=mul_invQ(U, G, psi, 1e-4);
\end{verbatim}}

The fourth argument is the required precision in the inversion. By
default it is $10^{-4}$.

Consider now the following program

{\footnotesize \begin{verbatim}
      // Program saved in FILE: p08.C
      #include "MDP_QCD.h"
      int main() {
        start();
        float precision=1e-4;
        gauge_field U;
        em_field G;
        fermi_field chi, psi, phi;
        U=hot();
        beta=6.0;
        multihit(U,1,20);
        G=compute_em_tensor(compute_plaquettes(U));

        kappa=0.125;
        c_SW=1.77;
        chi=new_fermi_config(U,G);

        psi=mul_Q(U,G,chi);
        phi=mul_invQ(U,G,psi,precision);
        phi=phi-chi;
        precision=abs(phi*phi)/(Nvol*Nc*Nc);
        printf("precision=%f\n",precision);
        stop();
        return 0;
      };
\end{verbatim}}

\noindent
It creates a gauge configuration $U$, a random fermionic
configuration $\chi$ and computes
\begin{eqnarray}
       \psi&=&Q[U]\chi \\
       \phi&=&Q^{-1}[U]\psi - \chi \\
       \delta&=&
	 \sum_{i,\alpha} \int d^4x \left[ \phi^i_\alpha(x) \right]^\ast
       \phi^i_\alpha (x)
\end{eqnarray}
It then checks whether $\delta$ is consistent with the input precision.
Note the command

{\footnotesize \begin{verbatim}
       chi=new_fermi_config(U,G);
\end{verbatim}}

\noindent
it returns a random fermi\_field $\chi$ with probability
\begin{equation}
       P[\chi]=\exp \left[-\chi^\dagger Q^\dagger[U] Q[U] \chi \right]
\end{equation}

Again, a {\tt fermi\_field psi} can be accessed as an $N_c \times 4$
matrix, {\tt psi(x)} or as complex number {\tt psi(x,i, alpha)}.

A more difficult task is the computation of the propagator of
eq.~(\ref{light}) from all points to all points.

\subsection{Stochastic light propagators}

The light propagator of eq.~(\ref{light}) can be computed, for each
gauge configuration, as an average \cite{stochastic1}  
\begin{equation} 
S_{\alpha\beta}^{ij}(x,y)[U]=\left\langle \varphi _\alpha 
^{\left[ k\right] i}(x)\left( \chi _\beta ^{\left[ k\right] j}(y)\right) 
^{*}\right\rangle _k  
\label{avg}
\end{equation} 
where for each $k$%
\begin{equation} 
\chi ^{\left[ k\right] }=Q[U]\varphi ^{\left[ k\right] }  
\end{equation} 

\noindent
The {\tt fermi\_field} configurations $\varphi ^{\left[ k\right]
}$ must be generated stochastically
\footnote{
The technique for generating these configurations has been recently
carried one step forward and combined with a ``maximal variance
reduction'' method to reduce the statistical noise \cite{michael}. 
This method is not implementd in {\tt MDP\_QCD}.}
 with probability  
\begin{equation} 
P[\varphi ]=\exp \left( -\varphi ^{\dagger }Q^{\dagger }[U]Q[U]\varphi 
\right)  
\end{equation} 
 
In {\tt MDP\_QCD} light propagators are objects of the class {\tt
light\_propagator}. A typical light propagators, $S$, is declared as

{\footnotesize \begin{verbatim}
       light_propagator S;
\end{verbatim}} 

Each {\tt light\_propagator S} contains two member arrays of {\tt %
fermi\_field} 

{\footnotesize \begin{verbatim} 
      fermi_field S.psi[Nfermi]; 
      fermi_field S.chi[Nfermi]; 
\end{verbatim}} 
 
These arrays can be created\footnote{%
Note that before generating or reading a {\tt light\_propagator} it is 
necessary to assign a value to the global variables {\tt kappa} and {\tt %
c\_SW} which are used by {\tt mul\_Q()}.} for a fixed gauge configuration, 
with the command: 

{\footnotesize \begin{verbatim} 
      S.stochastic_generation(U,G); 
\end{verbatim}} 

\noindent
where {\tt U} is the gauge field and {\tt G} if the electromagnetic
tensor. Here is how it is implemented

{\footnotesize \begin{verbatim} 
      light_propagator::stochastic_generation(gauge_field U, em_field G) {
        register site k;
        printf("Start stochastic generation: Nfermi=%i\n", Nfermi);
          for(k=0; k<Nfermi; k++) {
          printf("Step %i:",k);
          psi[k]=new_fermi_config(U,G,0.0001);
          chi[k]=mul_Q(U,G,psi[k]);
        };
        printf("Stochastic generation terminated.\n");
      };
\end{verbatim}} 

This code is the time expensive, therefore it may be a
good idea to save the fermionic configurations once they have been
generated. 

In the following program a stochastic propagator is created and saved
for the 60th of our gauge configurations.

{\footnotesize \begin{verbatim} 
      // Program saved in FILE: p09.C
      #include "MDP_QCD.h"
      int main() {
         start();
         gauge_field U;
         em_field G;
         light_propagator S;
         read(U, "gauge_config_beta_6.0", 60);
         kappa=0.1345; c_SW=1.77;
         G=compute_em_tensor(compute_plaquettes(U));
         S.stochastic_generation(U,G);
         write(S,"propagator.1345.177.", 60);
         stop();
         return 0;
      };
\end{verbatim}} 

There are two basic ways to extract informations from a {\tt
stochastic\_propagator S}. 
\begin{itemize}
\item
{\tt S(x,i,y,j,k)} returns the spin Matrix $N_{spin}\times N_{spin}$  
\begin{equation} 
\varphi _\alpha ^{\left[ k\right] i}(x)\left( \chi 
_\beta ^{\left[ k\right] j}(y)\right) ^{*}  
\end{equation}  
\item
{\tt S(x,i,y,j)} returns the same spin Matrix $N_{spin}\times
N_{spin}$ averaged on the spinorial configurations. 
\begin{equation} 
\left\langle \varphi _\alpha ^{\left[ k\right] 
i}(x)\left( \chi _\beta ^{\left[ k\right] j}(y)\right) ^{*}\right\rangle _k  
\end{equation} 
\end{itemize}

The following program computes, for just one gauge configuration, 
the two point correlation function of eq.~(\ref{2cp})
when $J(x)$ is the current $\overline{b}\gamma^5q$. Note that the
propagator for the $b$ is computed in the static approximation,
i.e. as product of links times $\frac12(1+\gamma^0)$. The light
propagator is read from a file.

{\footnotesize \begin{verbatim}
      // Program saved in FILE: p10.C
      #include "MDP_QCD.h"
      int main() {
        start();
        gauge_field U;
        em_field    G;
        light_propagator S;
        site x,y,z;
        int i,j,L=2;
        Matrix H(Nc,Nc);
        Matrix A;
        Complex result=0;
        read(U, "gauge_config_beta_6.0", 60);
        G=compute_em_tensor(compute_plaquettes(U));
        kappa=0.1345; c_SW=1.77; // They must correspond to the S saved!
        read(S,U,G,"propagator.1345.177.",60);
        x=0; y=move(x,0,L);
        H=1;
        A=Gamma5*0.5*(1+Gamma[0])*Gamma5
        for(z=x; z!=y; z=up[z][0]) H=H*hermitian(U(z,0));
        for(i=0; i<Nc; i++)
          for(j=0; j<Nc; j++)
            result=result+trace(S(y,i,x,j)*A)*H(j,i);
        printf("result=%f + %fI\n", real(result), imag(result));
        stop();
        return 0;
      };
\end{verbatim}}

Some care is necessary when simulating a process with more than one
light quark. The safe way is to define many stochastic propagators
corresponding to the different contractions. To save memory and time it
is possible to use just one {\tt light\_propagator} to compute
all of the ``physical'' propagators (assuming the corresponding quarks
are degenerate in mass). In this case the average of
eq.~(\ref{avg}) cannot be computed independently for each of the
propagators and unwanted correlations must be somehow eliminated.

\subsection{Smearing}

To improve the overlap between a current, $J^\dagger$, and the physical
state one wants to study, $| 1_J \rangle$, 
it is common to smear the light propagator. 
To this scope the gauge invariant ``Wupperthal'' 
smearing\cite{smearing} is implemented

{\footnotesize \begin{verbatim}
      float epsilon=0.25;
      int   num_iter=60;
      psi=smearing(chi,epsilon,num_iter);
\end{verbatim}}

\noindent
This smears the $\chi$ field with a smearing parameter $\epsilon$,
iterates the procedure {\tt num\_iter} times, then saves the smeared
field in $\psi$.

Here is how the function {\tt smearing} is implemented

{\footnotesize \begin{verbatim}
       fermi_field smearing(fermi_field psi, gauge_field U, 
                            float epsilon, int level) {
         fermi_field phi;
         int n;  
         register site x;
         printf("Smearing fermionic configuration...\n"); 
         for(n=0; n<level; n++) {
           if(n>0) psi=phi;
           for(x=0; x<Nvol; x++) 
             phi(x)=psi(x)+epsilon*
               (U(x,1)*psi(up[x][1])+
                U(x,2)*psi(up[x][2])+
                U(x,3)*psi(up[x][3])+
                hermitian(U(dw[x][1],1))*psi(dw[x][1])+
                hermitian(U(dw[x][2],2))*psi(dw[x][2])+
                hermitian(U(dw[x][3],3))*psi(dw[x][3]));		     
         };
         prepare(phi);
         return phi;
       };
\end{verbatim}}

To smear the sink (and/or the source) of a stochastic propagator it is
necessary to apply the smearing function to its member fields

{\footnotesize \begin{verbatim}
      for(n=0; n<Nfermi; n++) {
         S.psi[n]=smearing(S.psi[n],epsilon,num_iter); // smear sink
         S.chi[n]=smearing(S.chi[n],epsilon,num_iter); // smear source
      }
\end{verbatim}}

\subsection{Examples: fuzzing and cooling}

Fuzzing is a kind of gauge inveriant smearing that is not explicitely 
implemented in {\tt MDP\_QCD} but there are different ways of doing
it. It essentially consists of performing an ordinar smearing using a
a ``fuzzed'' gauge configuration instead of the normal one. Here is
how it could be implemented

{\footnotesize \begin{verbatim}
      // EXAMPLE CODE: function not implemented!
      fermi_field fuzzing(fermi_field psi, gauge_field U, int num_iter) {
        fermi_field phi;
        gauge_field Ufuzzed;
        site x;
        int mu,i;
        for(i=0; i<num_iter; i++)
          for(x=0; x<Nvol; x++)
            for(mu=0; mu<Ndim; mu++) {
              Ufuzzed(x,mu)=project_on_SUn(U(x,mu)+0.25*staple(U,x,mu));
          };
        phi=smearing(psi,Ufuzzed,0.25,2);
        prepare(phi);
        return phi;
      };
\end{verbatim}}

The function {\tt project\_on\_SUn} is not implemented in {\tt MDP\_QCD}
and there are different ways to do it. Its job is to return some kind
of projection of the matrix, which is passed as argument, on the
$SU(N_c)$ group. This procedure is called ``cooling''.
One way of implementing it is

{\footnotesize \begin{verbatim}
       // EXAMPLE CODE: function not implemented!
       Matrix project_on_SUn(Matrix A) {
         Matrix logA, B(Nc,Nc);
         int i;
         B=0;
         logA=log(A);         
         for(i=1; i<=Nadj; i++)
           B+=Generator[i]*real(0.5*trace(Generator[i]*logA));
         B=exp(B);
         prepare(B);
         return B;
       };
\end{verbatim}}

Note that {\tt Generator[i]} form a basis of $SU(N_c)$ and are built in
in {\tt MDP\_QCD}. This code is very slow, but it can be speeded up by
decreasing the precision in the computation of {\tt log()} and {\tt
exp()} for matrices. This is set in the flag {\tt PRECISION} declared in
the file {\tt MDP\_Lib1.h}. In any case this is usually not the critical
part of a lattice simulation.

\subsection{Basic algorithms: {\tt multihit()}}
 
The multihit code consists of the following iteractive 
procedure \cite{rothe}\cite{metropolis}
 
\begin{itemize} 
\item[$\blacktriangleright$]
For each link of the lattice 
 
\item[(1)]
Generate a random matrix 
 
\item[(2)]
Compute the variation in the action, $\delta S$ 
 
\item[(3)]  
If $\delta S>x$, where $x$ is random real number, substitute the new 
link to the old one 
 
\item[$\circlearrowright$]
Repeat steps $\{1,2,3\}$ {\tt n\_hits} times 
 
\item[$\circlearrowright$]
Repeat from $\blacktriangleright$ {\tt n\_iter} times 
\end{itemize} 
 
This algorithm is implemented in the following code

{\footnotesize \begin{verbatim} 
      void multihit(gauge_field U, int n_hit=10, int n_iter=1) {
         register int     i,hit,mu,nu; 
         register site    x;
         register double  delta_action, efficency;
         long    updated;
         Matrix  Unew,UUU;
         for(i=0; i<n_iter; i++) {
            printf("Multihit step n.%i, beta=%f...\n",i,beta);
            updated=0;
            for(x=0; x<Nvol; x++) {
               for(mu=0; mu<4; mu++) {
                  UUU=staple(U,x,mu);
                  for(hit=0; hit<n_hit; hit++) {
                     Unew=Random.SU(Nc)*U(x,mu);
                     delta_action=exp(beta/Nc*
                        real(trace(UUU*hermitian(Unew-U(x,mu)))));
     	             if(delta_action>Random.Float()) {
                        U(x,mu)=Unew;
                        updated++;
                        hit=n_hit;
                     };
                  };
               };
            };
            efficency=100.0*updated/(Nvol*Ndim);
            printf("... Efficency: %.2f%%\n", efficency);
         };
      };
\end{verbatim}} 

The parameter called {\tt efficiency}, 
computed by this function and printed on the standard output,
is the ratio between the number of links that has been updated
on each multi-hit step and the total number of links.
 
\subsection{Basic algorithms: {\tt mul\_Q()} and {\tt mul\_invQ()}}

The function {\tt mul\_Q()} performs multiplication by the fermionic 
matrix as it appears in the SW action. Here is a possible implementation:

{\footnotesize \begin{verbatim}
      fermi_field mul_Q(gauge_field U, em_field G, 
                        fermi_field psi, int sign=1) {
         // Slow Version!
         site x;
         int mu,nu;
         fermi_field tmp;
         tmp=psi;
         for(x=0; x<Nvol; x++)
            for(mu=0; mu<Ndim; mu++) {
               tmp(x)-=kappa*mul_left(1-sign*Gamma[mu],
                        U(x,mu)*psi(up[x][mu]))
	             +kappa*mul_left(1+sign*Gamma[mu], 
	                hermitian(U(dw[x][mu],mu))*psi(dw[x][mu]));
               for(nu=mu+1; nu<Ndim; nu++)
	          tmp(x)-=kappa*c_SW*left_mul(Sigma[mu][nu],
		        G(x,mu,nu)*psi(x));
            };
         prepare(tmp);
         return tmp;
      };
\end{verbatim}}

This would work but, in practice,
the real implementation has been optimized to reduce the
number of loops.

In the real implementation 
the multiplication when $c_{SW}\neq 0$ is only 80\% slower than 
when $c_{SW}=0$ (because the ``clover'' term is not computed). \\
The command

{\footnotesize \begin{verbatim} 
      psi=mul_invQ(U, G, chi, delta); 
\end{verbatim}} 

\noindent
returns $\psi=Q^{-1}[U]\chi$. The inversion is performed using 
the technique of minimal residue \cite{montvay}.
Note that the electromagnetic tensor {\tt G} is passed as an
argument as well.  This is redundant but it is much faster if many 
inversions have to be performed on the same gauge configuration
and $c_{SW}$ is different from zero.

The minimal residue scheme consists of the following iterative procedure
 
\begin{itemize} 
\item[$\blacktriangleright$]
  $\phi _0=\chi ,$ $r_0=\chi -Q[U]\chi $ 
 
\item[(1)]
  $\alpha _i=\frac{(Q[U]r_i)\cdot r_i}{(Q[U]r_i)\cdot (Q[U]r_i)}$ 
 
\item[(2)]
  $\phi _{i+1}=\phi _i+\alpha _ir_i,$ $r_{i+1}=r_i-\alpha _iQ[U]r_i$ 
 
\item[$\circlearrowright$]  
iterate step $\{1,2\}$ until $\left| r_i\right| ^2$ is less then the 
required precision ({\tt delta/(Nvol*Nc*Nc)}${\tt )}$. 

\item[$\blacksquare$]
$\phi _i$ converges towards $\psi$. 
\end{itemize}

\noindent 
Here is how it is implemented

{\footnotesize \begin{verbatim} 
      fermi_field mul_invQ(gauge_field U, em_field G, fermi_field chi,
                           float delta_res=0.0001, int sign=1) {
        printf("Inverting the Q matrix with kappa=%f, c_SW=%f ...\n",
                kappa, c_SW);
        fermi_field psi, res, Qres;
        float residue;
        Complex alpha;
        int step=0;
        psi=chi;
        res=chi-mul_Q(U,G,chi,sign);
        do {
          Qres=mul_Q(U,G,res,sign);
          alpha=(Qres*res)/(Qres*Qres);
          psi=psi+alpha*res;
          res=res-alpha*Qres;
          residue=real(res*res)/(Nvol*Nc*Nc);
          step++;
          printf("   Step %i =>%f\n", step, residue);
        } while (residue>delta_res);
        printf("... %i steps, precision reached=%e\n", step, residue);
        prepare(psi);
        return psi;
      };
\end{verbatim}} 

\subsection{Input/Output}

One gauge configuration {\tt U} can be saved with the command

{\footnotesize \begin{verbatim} 
      write(U, "filename", g); 
\end{verbatim}} 

\noindent 
where {\tt g} is an integer. Assuming for example that {\tt g} is 36,
the configuration is saved in a file called {\tt filename.00036}.
The command to load this file in a gauge configuration {\tt U} is

{\footnotesize \begin{verbatim} 
      read(U, "filename", g); 
\end{verbatim}} 

\noindent
Using properties of $SU(N_c)$ the configuration is compressed when it
is saved and automatically decompressed when it is read. After the
compression each link occupies $12 (N_c^2-N_c)$ bytes instead of
$16 N_c^2$.

Fermionic configurations are saved and loaded with the commands

{\footnotesize \begin{verbatim} 
      write(psi, "filename", g, n); 
      read(psi,  "filename", g, n); 
\end{verbatim}} 

\noindent 
to/from a file {\tt filename.}{\it g}{\tt .}{\it n}.

A {\tt light\_progator S} is saved with the command

{\footnotesize \begin{verbatim} 
      write(S, "filename", g); 
\end{verbatim}}

\noindent
which calls

{\footnotesize \begin{verbatim}
      for(n=0; n<Nfermi; n++) write(S.psi[n], "filename", g, n); 
\end{verbatim}} 

\noindent
It can be read with

{\footnotesize \begin{verbatim}
      G=compute_em_tensor(compute_plaquettes(U));
      read(S, U, G, "filename",g);
\end{verbatim}}

\noindent
Note that it is necessary to pass the gauge field and the
electromagnetic field to this function. They are necessary to rebuild
the {\tt S.chi[Nfermi]} fields from the {\tt S.psi[Nfermi]}. $U$ must
be exactly the same gauge configuration on which the {\tt
S.phi[Nfermi]} fields had been generated.

\subsection{Memory and speed}

\begin{table}
\begin{center}
\begin{tabular}{|l|l|}
\hline
object class & memory (bytes) \\ \hline
{\tt Matrix} & $11+rows \times columns$ \\
{\tt gauge\_field} & $5+32 \times N_{volume} \times N_c^2$ \\
{\tt pl\_field} & $5+96 \times N_{volume} \times N_c^2$ \\
{\tt em\_field} & $5+96 \times N_{volume} \times N_c^2$ \\
{\tt fermi\_field} & $5+32 \times N_{volume} \times N_c$ \\  
{\tt light\_propagator} & $(5+32 \times N_{volume} \times
N_c)\times 2 \times N_{fermi}$ \\ \hline
\end{tabular}
\end{center}
\caption{Memory usage for the different fields. $N_{volume}$ is
for the total number of lattice sites.}
\label{mem}
\end{table}

Tab.~(\ref{mem}) shows the required memory to store 
each of the fundamental objects (as function of the lattice size 
defined in {\tt MDP\_settings.h}).

The function {\tt mul\_invQ()} allocates 3 {\tt fermi\_field} and one
of them is returned.
\begin{table}
\begin{center}
\begin{tabular}{|c|lllll|}
\hline 
lattice  & $U(1)$ & $SU(2)$ & $SU(3)$ & $SU(4)$ & $SU(5)$ \\ \hline
$2^3 \times 3$  & 0.03  & 0.03  & 0.10 & 0.33  & 0.47\\
$4^3 \times 6$  & 0.28  & 0.50  & 1.21 & 2.72  & 5.99 \\
$6^3 \times 9$  & 1.25  & 2.65  & 6.18 & 14.54 & 32.94 \\
$8^3 \times 12$ & 3.84  & 8.02  & 20.07 & 45.18& 93.65 \\ 
$12^3 \times 18$& 20.14 & 38.79 & 98.07 & -    &  - \\\hline
\end{tabular}
\end{center}
\caption{Time in seconds to perfrom one multihit step (1 hit per link)
on a SUN UltraSPARC 5.}
\label{figure2}
\end{table}

\begin{table}
\begin{center}
\begin{tabular}{|c|lllll|}
\hline 
lattice  & $U(1)$ & $SU(2)$ & $SU(3)$ & $SU(4)$ & $SU(5)$ \\ \hline
$2^3 \times 3$ & 0.05   & 0.07  & 0.13 & 0.24  & 0.43  \\
$4^3 \times 6$ & 0.46   & 1.08  & 2.20 & 4.19  & 6.36  \\
$6^3 \times 9$ & 2.24   & 5.34  & 11.20& 19.37 & 31.95 \\
$8^3 \times 12$ & 7.14  & 14.87 & 35.80& 62.24 & 105.00\\ 
$12^3 \times 18$& 35.74 & 84.32 & 174.90 & -    &  -   \\ \hline
\end{tabular}
\end{center}
\caption{Time in seconds to generate one random fermionic configuration (it
involves one call to {\tt mul\_invQ()} with a precision of $10^{-4}$),
far from the chiral limit. Computed on a SUN UltraSPARC 5.}
\label{figure3}
\end{table}

All the other functions have been optimized and essentially they 
do not use more memory than that allocated to their arguments plus
the memory required to store the argument that is returned. It is important
to remeber that when a field is returned it is NOT copied, but only
the pointer to the physical memory is
copied. In other words the memory occupied by a field object that has
to be returned by a function is the same memory where the field will
be stored after the function has returned.

Since the time required to copy memory has been reduced to its minimum
any function declared in {\tt MDP\_QCD.h} is very fast.
{\tt MDP\_QCD} enables to run significative lattice simulations on a
SUN UltraSPARC or on a Pentium II PC.
Some benchmarks executed are reported in
tab.~(\ref{figure2}) and tab.~(\ref{figure3}) for the most 
common operations.
\vskip 5mm
It is important to stress that {\tt MDP\_QCD} does not include any
optimization for parallel machines and many of the trick that are
implemented may result in a slowing down on parallel computers without
shared memory. \\
Internally all the fields and matrices are seen as one dimensional
arrays of {\tt Complex}. This gives some benefit on vectorial machines
providing the compiler is able to optimize it.
\newpage

\appendix

\section{Lattice notation and conventions}

\begin{itemize}
\item  Equivalence between Lattice Euclidean and Minkowsky Space
quantities (where $a$ is the lattice spacing). 
\begin{equation}
\begin{tabular}{|c|c|}
\hline
Euclidean Lattice & Minkowsky Space \\ \hline
$x^0 $ & $ i a^{-1} x^0$ \\
$x^i $ & $ a^{-1} x^i$ \\ \hline
$\partial^4 $ & $ -i a \partial _0$ \\ 
$\partial^i $ & $a \partial _i$ \\ \hline
$A^0 $ & $-i a A_0$ \\
$A^i $ & $a A_i$ \\ \hline
$F^{0i} $ & $-i a^2 F_{0i}$ \\
$F^{ij} $ & $ a^2 F_{ij}$ \\ \hline
$\gamma^0 $ & $\gamma ^0$ \\
$\gamma^i $ & $-i \gamma ^i$ \\ 
$\gamma^5 $ & $\gamma ^5$  \\ \hline
\end{tabular}
\end{equation}
\item  Metric 
\begin{equation}
{g}^{\mu \nu }=-\delta ^{\mu \nu }=\text{diag}(-1,-1,-1,-1)
\end{equation}

\item  Pauli matrices 
\begin{eqnarray}
\texttt{sigma[1]} &=& \sigma _1=\left( 
\begin{array}{ll}
0 & 1 \\ 
1 & 0
\end{array}
\right) \\
\texttt{sigma[2]} &=& \sigma _2=\left( 
\begin{array}{ll}
0 & -i \\ 
i & 0
\end{array}
\right) \\
\texttt{sigma[3]} &=& \sigma _3=\left( 
\begin{array}{ll}
1 & 0 \\ 
0 & -1
\end{array}
\right) 
\end{eqnarray}
they undergo the commutation relation 
\begin{equation}
\lbrack \sigma _i,\sigma _j]=2i\varepsilon ^{ijk}\sigma _k
\end{equation}

\item  Dirac matrices in 
Dirac representation (the same convention of \cite{rothe} and UKQCD)
\begin{eqnarray}
\texttt{Gamma[0]} & = &
\gamma^0=\left( 
\begin{array}{ll}
{\bf 1} & 0 \\ 
0 & -{\bf 1}
\end{array}
\right) \\
\texttt{Gamma[i]} & =&  
\gamma^i=\left(  
\begin{array}{ll}
0 & -i\sigma _i \\ 
i\sigma _i & 0
\end{array}
\right) \\
\texttt{Gamma5} & = &
\gamma^5=\left( 
\begin{array}{ll}
0 & {\bf 1} \\ 
{\bf 1} & 0
\end{array}
\right)   \\
\texttt{Gamma1} &=& \delta^{\mu \nu } 
=\frac 12\{\gamma^\mu ,\gamma^\nu \} \\
\texttt{Sigma[mu][nu]} &=& \sigma^{\mu \nu } 
=\frac i2[\gamma^\mu ,\gamma^\nu ] \\
\texttt{Pleft} & = & P_{left}=\frac{1-\gamma^5}2 \\
\texttt{Pright} & = & P_{right}=\frac{1+\gamma^5}2
\end{eqnarray}
Note that all the $\gamma$ and $\sigma$ matrices are hermitian and, by
definition, $\gamma^5=\gamma^0 \gamma^1 \gamma^2 \gamma^3$.

\item  Traces 
\begin{eqnarray}
tr(\gamma^\mu \gamma^\nu ) &=& 4\delta ^{\mu \nu } \\
tr(\gamma^\mu \gamma^\nu \gamma^\rho ) &=&0      \\
tr(\gamma^\mu \gamma^\nu \gamma^\rho 
\gamma^\sigma ) &=&4(\delta^{\mu \nu }\delta^{\rho
\sigma }-\delta^{\mu \rho }\delta^{\nu \sigma }+\delta^{\mu
\sigma }\delta^{\rho \nu }) \\
tr(\gamma^5 \gamma^\mu \gamma^\nu 
\gamma^\rho \gamma^\sigma ) &=&4\epsilon^{\mu \nu \rho
\sigma }
\end{eqnarray}
where $\epsilon ^{0123}=\epsilon^{0123}=-1.$

\item Gell-Mann matrices, $\texttt{Lambda[i]}=\lambda_i$
\begin{eqnarray}
\lambda ^1 &=&\left( 
\begin{array}{lll}
0 & 1 & 0 \\ 
1 & 0 & 0 \\ 
0 & 0 & 0
\end{array}
\right) \qquad \lambda ^2=\left( 
\begin{array}{lll}
0 & -i & 0 \\ 
i & 0 & 0 \\ 
0 & 0 & 0
\end{array}
\right)  \\
\lambda ^3 &=&\left( 
\begin{array}{lll}
1 & 0 & 0 \\ 
0 & -1 & 0 \\ 
0 & 0 & 0
\end{array}
\right) \qquad \lambda ^4=\left( 
\begin{array}{lll}
0 & 0 & 1 \\ 
0 & 0 & 0 \\ 
1 & 0 & 0
\end{array}
\right)  \\
\lambda ^5 &=&\left( 
\begin{array}{lll}
0 & 0 & -i \\ 
0 & 0 & 0 \\ 
i & 0 & 0
\end{array}
\right) \qquad \lambda ^6=\left( 
\begin{array}{lll}
0 & 0 & 0 \\ 
0 & 0 & 1 \\ 
0 & 1 & 0
\end{array}
\right)  \\
\lambda ^7 &=&\left( 
\begin{array}{lll}
0 & 0 & 0 \\ 
0 & 0 & -i \\ 
0 & i & 0
\end{array}
\right) \qquad \lambda ^8=\frac 1{\sqrt{3}}\left( 
\begin{array}{lll}
1 & 0 & 0 \\ 
0 & 1 & 0 \\ 
0 & 0 & -2
\end{array}
\right) 
\end{eqnarray}

\item Discrete symmetries\cite{bernard}:
\begin{eqnarray}
P: S_{\alpha \beta}^{ij}(x,y)[U]&=&
\gamma^0_{\alpha\alpha'} S_{\alpha' \beta'}^{ij}(x^P,y^P)[U^P]
\gamma^0_{\beta'\beta} \\ 
C: S_{\alpha \beta}^{ij}(x,y)[U]&=&
(\gamma^0\gamma^2)_{\alpha\alpha'} S_{\alpha' \beta'}^{ji}(y,x)[U^C]
(\gamma^2\gamma^0)_{\beta'\beta} \\ 
T: S_{\alpha \beta}^{ij}(x,y)[U]&=&
(\gamma^0\gamma^5)_{\alpha\alpha'} S_{\alpha' \beta'}^{ij}(x^T,y^T)[U^T]
(\gamma^5\gamma^0)_{\beta'\beta} \\ 
H: S_{\alpha \beta}^{ij}(x,y)[U]&=&
\gamma^5_{\alpha\alpha'} S_{\alpha' \beta'}^{ji}(y,x)[U]
\gamma^5_{\beta'\beta} 
\end{eqnarray}
$U^P,U^C,U^T$ are the 
parity reversed, charge conjugate, time reversed gauge configuration
respectively.
\end{itemize}

\section{Basic syntax for {\tt MDP\_QCD}}

\begin{itemize}
\item
Classes defined in the file {\tt MDP\_Lib1.h}
included by {\tt MDP\_QCD.h:} 

{\footnotesize \begin{verbatim} 
     class Toy_class; 
     class Matrix; 
     class Random_generator; 
     class JackBoot; 
\end{verbatim}} 
\item Classes defined in the file {\tt MDP\_QCD.h:} 

{\footnotesize \begin{verbatim} 
     class gauge_field public: Toy_class<Complex, Nvol*Ndim*Nc*Nc>; 
     class pl_field    public: Toy_class<Complex, Nvol*6*Nc*Nc>; 
     class em_field    public: Toy_class<Complex, Nvol*6*Nc*Nc>; 
     class fermi_field public: Toy_class<Complex, Nvol*Nc*Nspin>; 
     class light_propagator; 
\end{verbatim}} 
The elements of each field-type class can be accessed as matrices

{\footnotesize \begin{verbatim}
     Matrix gauge_field::operator() (site x, int mu);
     Matrix em_field::operator()    (site x, int mu, int nu);
     Matrix pl_field::operator()    (site x, int mu, int nu);
     Matrix fermi_field::operator() (site x);
\end{verbatim}}
or as complex numbers

{\footnotesize \begin{verbatim}
     Complex gauge_field::operator() (site x, int mu, int i, int j);
     Complex em_field::operator()    (site x, int mu, int nu, 
                                      int i,int j);
     Complex pl_field::operator()    (site x, int mu, int nu,
                                      int i, int j);
     Complex fermi_field::operator() (site x, int i, int alpha);
\end{verbatim}}

To save memory {\tt em\_field} and {\tt pl\_field} only store those
with components {\tt mu}$<${\tt nu}.
The class {\tt light propagator} has peculiar class members and they
are discussed in the proper section.

\item Global functions: 

{\footnotesize \begin{verbatim} 
     site move(site x, int mu, int lmu); 
     site move(site x, int mu, int lmu, int nu, int lu); 
     site coordinate_space(x); 
     site position(int x0, int x1, int x2, int x3); 
     site position(int x0, site x_space); 
     float dist(site x, site y); 
     void start(); 
     void stop(); 
     Matrix plaquette(gauge_field U, site x, int mu, int nu); 
     Matrix staple_up(gauge_field U, site x, int mu, int nu); 
     Matrix staple_dw(gauge_field U, site x, int mu, int nu); 
     Matrix staple(gauge_field U, site x, int mu, int nu); 
     float average_plaquette(gauge_field U); 
     gauge_field cold(); 
     gauge_field hot(); 
     float  action(gauge_field U); 
     void   multihit(gauge_field U, int n_steps, n_hits); 
     pl_field compute_plaquettes(gauge_field U);
     em_field compute_em_tensor(pl_field Up);
     fermi_field mul_Q(gauge_field U, em_field T, 
                       fermi_field psi, int dagger=1); 
     fermi_field mul_invQ(gauge_field U, em_field T,
                          fermi_field psi, 
                          float delta=1e-4, int dagger=1); 
     fermi_field new_fermi_config(gauge_field U, em_field T); 
     fermi_field smearing(fermi_field psi, gauge_field U,
                          float epsilon, int level);
\end{verbatim}} 
 
\item Global functions for Input/Output: 

{\footnotesize \begin{verbatim} 
      void write(gauge_field U, char filename[], int gauge); 
      void  read(gauge_field U, char filename[], int gauge); 
      void write(fermi_field psi, char filename[], int gauge, int nfermi); 
      void  read(fermi_field psi, char filename[], int gauge, int nfermi); 
      void write(light_ptopagator S, char filename[], int gauge); 
      void  read(light_propagator S, gauge_field U, em_field T,
                 char filename[], int gauge); 
\end{verbatim}} 
 
\item Global constants (defined in {\tt MDP\_Settings.h}, included by {\tt
MDP\_QCD.h}). They can be modified by the programmer.
 
\begin{equation} 
\begin{tabular}{|l|l|l|} 
\hline 
& {\tt C++ with MDP\_QCD} & Value \\ \hline 
lattice sites in time & {\tt Nx0} & {\tt 8} \\
lattice sites is $x_1$ & {\tt Nx1} & {\tt 4} \\
lattice sites is $x_2$ & {\tt Nx2} & {\tt 4} \\
lattice sites is $x_3$ & {\tt Nx3} & {\tt 4} \\
n${{}^{\circ }}$ colors & {\tt Nc} & {\tt 3} \\
n${{}^{\circ }}$ fermi configs & {\tt Nfermi} & {\tt 10} \\ \hline 
\end{tabular} 
\end{equation} 
 
\item Global constants that should not be modified by the
programmer and can be used in {\tt main{}} and in functions.
 
\begin{equation} 
\begin{tabular}{|l|l|l|} 
\hline 
& {\tt C++ with MDP\_QCD} & Value \\ \hline 
lattice sites in space & {\tt Nspace} & {\tt Nx1*Nx2*Nx3} \\
lattice sites in total & {\tt Nvol} & {\tt Nx0*Nspace} \\ 
n${{}^{\circ }}$ dimensions & {\tt Ndim} & {\tt 4} \\ 
n${{}^{\circ }}$ spin components & {\tt Nspin} & {\tt 4} \\
elements in adjoint & {\tt Nadj} & {\tt Nc*Nc-1} \\ 
& {\tt DAGGER} & {\tt -1} \\ 
new type & {\tt site} & {\tt long} \\
logical operator & {\tt and} & {\tt \&\&} \\
logical operator & {\tt or} & {\tt ||} \\  \hline  
\end{tabular} 
\end{equation} 

\item Global variables: 

\begin{equation} 
\begin{tabular}{|l|l|l|} 
\hline 
& {\tt C++ with MDP\_QCD} & type \\ \hline 
$\beta $ & {\tt beta} & {\tt float} \\ 
$\kappa $ & {\tt kappa} & {\tt float} \\
$c_{SW}$ & {\tt c\_SW} & {\tt float} \\ 
$x+\widehat{\mu }$ & {\tt up[x][mu]} & {\tt site} \\
$x-\widehat{\mu }$ & {\tt dw[x][mu]} & {\tt site} \\
$x_\mu $ & {\tt co[x][mu]} & {\tt int} \\ 
$\gamma _{\mu =0..3}$ & {\tt Gamma[mu]} & {\tt Matrix }$4\times 4$ \\
$\gamma ^5$ & {\tt Gamma5} & {\tt Matrix }$4\times 4$ \\ 
$\sigma _{\mu \nu }$ & {\tt Sigma[mu][nu]} & {\tt Matrix }$4\times 4$ \\   
$\sigma ^{i=0..3}$ ($\sigma ^0=1$) & {\tt sigma[i]} & {\tt Matrix }$2\times 
2 $ \\ 
$\lambda ^{a=0..8}$ ($\lambda ^0=1$) & {\tt Lambda[a]} & {\tt Matrix }$%
3\times 3$ \\
$\omega ^{c=0..N_c^2}$ ($\omega ^0=1$) & {\tt Generator[c]} &  
{\tt Matrix }$N_c\times N_c$ \\ \hline 
\end{tabular} 
\label{whattable}
\end{equation}
The variables $\beta$, $\kappa$ and $c_{SW}$ must be initialized
in the main program before they are used.
Note that the matrices $\omega ^{c=1..N_c^2}$ form a general basis 
of Hermitian  matrices for $SU(N_c)$. They coincide with the Pauli 
matrices for $N_c=2$, but not with the Gell-Mann matrices for $N_c=3$. 
\end{itemize}
 
\newpage
\section{{\tt Toy\_class} and how it works}

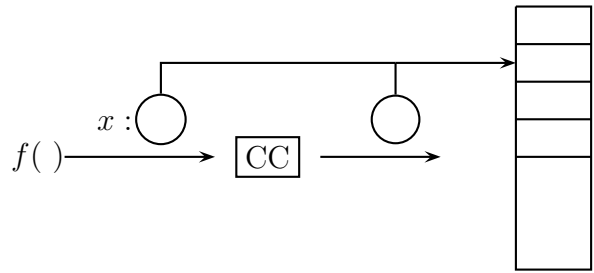
\begin{figure}[tbp] 
\begin{center} 
\begin{pspicture}(1,0)(10,4.5) 
\rput[cr]{0}(2,2){$f (\ )$} 
\rput[c]{0}(3,2.5){$x:\:$\circlenode{x}{{\white{}R}}} 
\rput[c]{0}(6.4,2.5){\circlenode{y}{{\white{}F}}} 
\rput[c]{0}(4.7,2){\psframebox{CC}} 
\pnode(8,3.25){muno} 
\pnode(6.4,3.25){mdue} 
\psline{->}(2,2)(4,2) 
\psline{->}(5.4,2)(7,2) 
\psline(8,4)(9,4)(9,0.5)(8,0.5)(8,4) 
\psline(8,3.5)(9,3.5) 
\psline(8,3)(9,3) 
\psline(8,2.5)(9,2.5) 
\psline(8,2)(9,2) 
\ncangle[angleA=90,angleB=180]{x}{mdue} 
\ncline{->}{mdue}{muno} 
\ncline{y}{mdue} 
\end{pspicture} 
\end{center} 
\par 
\smallskip \vskip 3mm \hrule \vskip 3mm \smallskip 
\caption{When the object $x$ is returned the copy constructor (CC) is called 
and it creates a temporary object. The objects are represented as circles. } 
\label{fig1b} 
\end{figure}

Some books\cite{core} about {\tt C++} warn the reader about the 
problems of returning objects. In fact when an object is returned by a 
function, it is copied by the copy constructor into a temporary object. When 
the function terminates the temporary object is returned (fig.~\ref{fig1b}).  
\begin{figure}[tbp] 
\begin{center} 
\begin{pspicture}(1,0)(10,4.5) 
\psframe[fillstyle=solid,fillcolor=gray](8,3)(9,3.5) 
\rput[cr]{0}(2,2){$f (\ )$} 
\rput[c]{0}(3,2.5){$x:\:$\circlenode[fillstyle=solid,fillcolor=gray]{x}{{\gray{}R}}} 
\rput[c]{0}(6.4,2.5){\circlenode{y}{{\white{}F}}} 
\rput[c]{0}(4.7,2){\psframebox{CC}} 
\pnode(8,3.25){muno} 
\pnode(6.4,3.25){mdue} 
\rput[c]{0}(7,1){$g(y)$} 
\psline{->}(2,2)(4,2) 
\psline(5.4,2)(7,2) 
\psline{->}(7,2)(7,1.3) 
\psline(8,4)(9,4)(9,0.5)(8,0.5)(8,4) 
\psline(8,3.5)(9,3.5) 
\psline(8,3)(9,3) 
\psline(8,2.5)(9,2.5) 
\psline(8,2)(9,2) 
\ncangle[angleA=90,angleB=180]{x}{mdue} 
\ncline{->}{mdue}{muno} 
\ncline{y}{mdue} 
\end{pspicture} 
\end{center} 
\par 
\smallskip \vskip 3mm \hrule \vskip 3mm \smallskip 
\caption{The temporary object created by the copy constructor (CC) is 
assigned to the argument of the calling function, then the original object $%
x $ is destroyed and the memory is deallocated.} 
\label{fig1c} 
\end{figure}
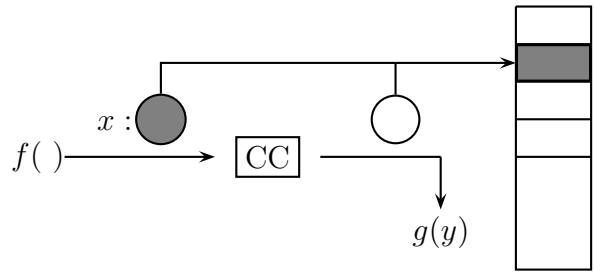 
 
If the original object to be returned contains a dynamically allocated 
pointer ({\tt p}) then its value is copied into the temporary object but the 
memory pointed by {\tt p} is not. 
The problem arises when the function terminates 
because the destructor of the original object is called and the memory 
pointed to by {\tt p} is deallocated (fig.~\ref{fig1c}). Hence the temporary 
object contains a pointer to a location of memory which is no longer 
allocated. 
 
When this situation occurs, there is sometimes a runtime error such as ``bus 
error" but often the program continues to run accessing to locations of 
memory containing random data without protection. 
 
Consider for example the following code: 

{\footnotesize \begin{verbatim} 
     // THIS CODE IS WRONG!!! 
     template <class T, long imax> class Toy_class { 
     public: 
        T *m; 
        Toy_class() { 
           m=new T[imax]; 
           for(i=0; i<imax; i++) m[i]=0; 
        }; 
      }; 
      Toy_class f() {                  // function f() 
         Toy_class<int,10> x; 
         x.p[3]=5; 
         return x; 
      }; 
      T g(Toy_class<int,10> y) {       // function g() 
         return y.p[3]; 
      }; 
      int main() {                     // main program 
         cout << g(f()); 
         return 0; 
      }; 
  
\end{verbatim}} 
 
It is supposed to print `5' on the screen, but it may give any random 
result: it is not safe! 
 
There are two standard ways to overcome this problem: 
 
\begin{itemize} 
\item  Redefine the copy constructor so that it will allocate new memory and 
will copy {\tt p[i]} for every {\tt i}. 

{\footnotesize \begin{verbatim} 
     Toy_class (Toy_class &x) { 
        if(p!=0) delete[] p; 
        p=new T[imax]; 
        for(int i=0; i<imax; i++) p[i]=x.p[i]; 
     }; 
\end{verbatim}} 
 
This is pretty safe, but it may force you to copy a huge amount of memory 
when it is not necessary (fig.~\ref{fig1d}). 
 
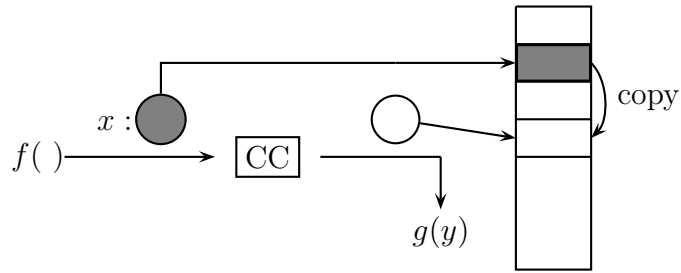
\begin{figure}[tbp] 
\begin{center} 
\begin{pspicture}(1,0)(10,4.5) 
\psframe[fillstyle=solid,fillcolor=gray](8,3)(9,3.5) 
\rput[cr]{0}(2,2){$f (\ )$} 
\rput[c]{0}(3,2.5){$x:\:$\circlenode[fillstyle=solid,fillcolor=gray]{x}{{\gray{}R}}} 
\rput[c]{0}(6.4,2.5){\circlenode{y}{{\white{}F}}} 
\rput[c]{0}(4.7,2){\psframebox{CC}} 
\pnode(8,3.25){muno} 
\pnode(6.4,3.25){mdue} 
\pnode(9,3.25){munor} 
\pnode(8,2.25){lastl} 
\pnode(9,2.25){lastr} 
\rput[c]{0}(7,1){$g(y)$} 
\psline{->}(2,2)(4,2) 
\psline(5.4,2)(7,2) 
\psline{->}(7,2)(7,1.3) 
\psline(8,4)(9,4)(9,0.5)(8,0.5)(8,4) 
\psline(8,3.5)(9,3.5) 
\psline(8,3)(9,3) 
\psline(8,2.5)(9,2.5) 
\psline(8,2)(9,2) 
\ncangle[angleA=90,angleB=180]{x}{mdue} 
\ncline{->}{mdue}{muno} 
\ncline{->}{y}{lastl} 
\nccurve[angleA=315,angleB=45]{->}{munor}{lastr}\Aput{copy} 
\end{pspicture} 
\end{center} 
\par 
\smallskip \vskip 3mm \hrule \vskip 3mm \smallskip 
\caption{A possible solution: The memory is copied by the copy constructor 
before the destructor of $x$ is called. } 
\label{fig1d} 
\end{figure} 
 
\item  Write all the functions so that their arguments are passed and 
returned by reference (so that the copy constructor is never called). Write 
the destructor in such a way that, when an object is destroyed, the memory 
pointed to by {\tt p} is not deallocated. 
It is also necessary to keep trace of 
the allocated memory and deallocate it somewhere in the program. This works, 
but only in very simple cases. This technique will become intractable 
in the presence of recursive functions which pass and return objects. 
\end{itemize} 
 
In this appendix a solution to this problem is proposed, based on the idea of 
attaching a {\tt FLAG} to each object. The {\tt FLAG} will contain 
information about the status of the object and it will be used to implement, 
in the most general framework, a simple and efficient way of passing and 
returning objects of any kind, even in recursive structures. Moreover all 
the tricks due to the dynamical allocations will be hidden within the basic 
methods of the class (constructor, copy constructor, destructor and 
assignment operator). 
 
The word ``efficient" here means that the program will automatically take 
care of the memory used and nothing will be copied if it is not necessary, 
hence the code will be optimized both in speed and memory usage. 
\footnote{ 
The technique that will be explained is used to implement the class {\tt 
Matrix}, {\tt gauge\_field}, {\tt pl\_field}, {\tt em\_field} 
and {\tt fermi\_field.}}
 
Suppose that one attaches a {\tt FLAG}, i.e. a new member variable, to each 
object 

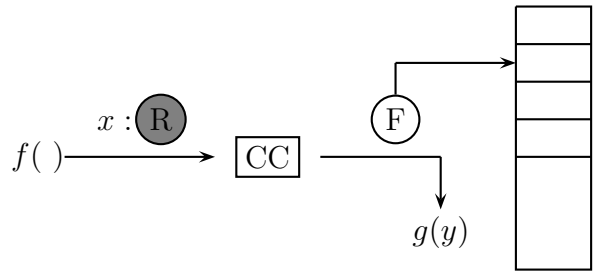
\begin{figure}[tbp] 
\begin{center} 
\begin{pspicture}(1,0)(10,4.5) 
\rput[cr]{0}(2,2){$f (\ )$} 
\rput[c]{0}(3,2.5){$x:\:$\circlenode[fillstyle=solid,fillcolor=gray]{x}{R}} 
\rput[c]{0}(6.4,2.5){\circlenode{y}{F}} 
\rput[c]{0}(4.7,2){\psframebox{CC}} 
\pnode(8,3.25){muno} 
\pnode(6.4,3.25){mdue} 
\rput[c]{0}(7,1){$g(y)$} 
\psline{->}(2,2)(4,2) 
\psline(5.4,2)(7,2) 
\psline{->}(7,2)(7,1.3) 
\psline(8,4)(9,4)(9,0.5)(8,0.5)(8,4) 
\psline(8,3.5)(9,3.5) 
\psline(8,3)(9,3) 
\psline(8,2.5)(9,2.5) 
\psline(8,2)(9,2) 
\ncline{->}{mdue}{muno} 
\ncline{y}{mdue} 
\end{pspicture} 
\end{center} 
\par 
\smallskip \vskip 3mm \hrule \vskip 3mm \smallskip 
\caption{The destructor of $x$ knows from the flag R that this object is 
going to be returned and it does not deallocate the memory. The temporary 
object $y$ carries an F flag, hence it will be deallocated later.} 
\label{fig1e} 
\end{figure}

{\footnotesize \begin{verbatim} 
     enum value {F, T, C, H} FLAG; 
\end{verbatim}} 
 
and one uses it in the following way: 
 
The {\tt FLAG} is always set to (F)REE by the constructor. If the object is 
not going to be returned its {\tt FLAG} will remain unchanged and the 
destructor will deallocate the memory pointed by its pointer when it is 
called. If the object is going to be returned its {\tt FLAG} is changed to 
(R)ETURN and its destructor will not deallocate the memory in this case. On 
the other side the copy constructor that is going to create the temporary 
object will check the {\tt FLAG} of its argument and if it is R, it will set 
the {\tt FLAG} of the temporary object to F (fig.~\ref{fig1e}). 
 
Moreover it is necessary to take care of the possibility of passing an F 
object as argument of a function: in this case the copy constructor will 
generate a temporary object with a C {\tt FLAG} and the destructor will be 
implemented in such a way that the memory pointed by a C will never be 
deallocated. 
 
\begin{figure}[t] 
\begin{center} 
\begin{pspicture}(0,0)(11,7) 
\psframe(.5,.5)(5,6.5) 
\psframe(.8,.8)(4.7,5.5) 
\psframe(1.1,2.3)(4.4,4) 
\rput[c]{0}(1.8,6){\circlenode{xa}{F}} 
\rput[c]{0}(1.8,4.5){\circlenode{xb}{C}} 
\rput[c]{0}(2.8,4.5){\circlenode{ya}{F}} 
\rput[c]{0}(1.8,3){\circlenode{xc}{C}} 
\rput[c]{0}(2.8,3){\circlenode{yb}{C}} 
\rput[c]{0}(3.8,3){\circlenode{za}{R}} 
\rput[c]{0}(3.8,1.3){\circlenode{zb}{F}} 
\ncline{->}{xa}{xb} 
\ncline{->}{xb}{xc} 
\ncline{->}{ya}{yb} 
\ncline{->}{za}{zb} 
\psframe(6.5,.5)(11,6.5) 
\psframe(6.8,.8)(10.7,3.3) 
\psframe(8.8,3.3)(10.4,5) 
\rput[c]{0}(7.8,6){\circlenode{la}{F}} 
\rput[c]{0}(7.8,2.3){\circlenode{lb}{C}} 
\rput[c]{0}(9.8,3.8){\circlenode{na}{R}} 
\rput[c]{0}(9.8,2.3){\circlenode{nb}{F}} 
\ncline{->}{la}{lb} 
\ncline{->}{na}{nb} 
\end{pspicture} 
\end{center} 
\par 
\smallskip \vskip 3mm \hrule \vskip 3mm \smallskip 
\caption{Different kinds of objects that may be passed to (and returned by) 
a function are represented. In these diagrams the copy constructor is 
automatically called when the arrow crosses the boundary of a box 
(representing a functional level in the program). } 
\label{fig2} 
\end{figure}
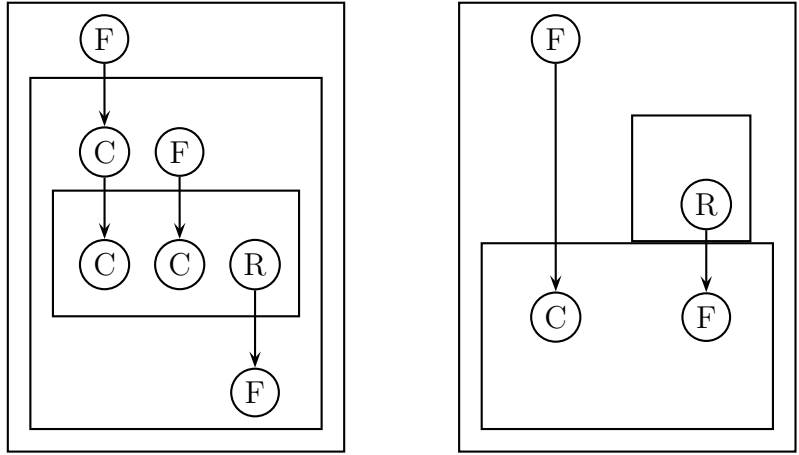

In other words a F object is a normal object that sooner or later is going 
to be destroyed (in the same function, or method, where it has been 
created), while a C one is a copy. Its pointer is pointing to memory 
allocated by someone else (a F object existing at an higher level). On the 
other side an object R exists only in the brief instant between the moment 
when the copy constructor is called and when its own destructor is called. 
The memory that it is pointing to is not deallocated because the new F 
temporary object created by the copy constructor will contain a reference to 
it. Such a memory will be deallocated later when the destructor of the 
temporary object will be called somewhere automatically. The three possible 
situations are illustrated in fig.~\ref{fig2} (left). 
 
To make the {\tt FLAG} usage even safer it is better to define it as a 
private member and define a function ({\tt prepare()}) that 
sets the {\tt FLAG} of the object to R before it is returned. 
 
A {\tt FLAG} value (H)yperlink allows an object to contain a pointer to an 
address of memory allocated by an object of a different class. So that
this memory will never be deallocated by the destuctor of the object.
 
According with these prescriptions the program class Toy\_class should be 
defined in the following way: 

{\footnotesize \begin{verbatim} 
      template <class T, long imax>
      class Toy_class {
      public:
        enum value {F,R,C,H} FLAG;
        T *m;  
        Toy_class() {
          register long i;
          FLAG=F;
          m=new T[imax];
          for(i=0; i<imax; i++) m[i]=0;
        };
        Toy_class(T *m0) {
          FLAG=H;
          m=m0;
        };
        Toy_class(const Toy_class &a) {
          m=a.m;
          switch(a.FLAG) {
          case F: FLAG=C; break;
          case R: FLAG=F; break;
          case C: FLAG=C; break;
          case H: FLAG=H; break;
          };
        };   
        ~Toy_class() {
          if((FLAG==F) && (m!=0)) delete[] m;
        };
        operator= (Toy_class a) {
          register long i;
          switch(FLAG) {
          case H:       
            for(i=0; i<imax; i++) m[i]=a.m[i];
            break;
          case F:
            if(a.FLAG==F) {
              if ((m!=0) && (m!=a.m)) delete[] m;
              m=a.m;
              a.m=0;
            } else {
              for(i=0; i<imax; i++) m[i]=a.m[i];
            };
            break;
          case C: 
            if(a.FLAG==F) {
              m=a.m;
              a.m=0;;
            } else {
              m=new T[imax];
              for(i=0; i<imax; i++) m[i]=a.m[i];
            };
            FLAG=F;
            break;
          case R: 
            error("What the Hell are you doing?"); 
            break;
          };
        };
        inline friend void prepare (Toy_class &a) {
          register long i;
          T* m;
          switch(a.FLAG) {
          case F: a.FLAG=R; break;
          case R: 
            printf("You shoud not call prepare(); twice\n"); 
            break;
          default:
            m=new T[imax];
            for(i=0; i<imax; i++) m[i]=a.m[i];
            a.m=m;
            a.FLAG=R;
            break;
          };
        };
      };
\end{verbatim}} 
 
Now the following program works! 

{\footnotesize \begin{verbatim} 
      Toy_class f() {               // function f() 
         Toy_class<int,10> x; 
         x.p[3]=5; 
         return x; 
      }; 
      T g(Toy_class<int,10> y) {    // function g() 
         return y.p[3]; 
      }; 
      int main() {                  // main program 
         cout << g(f()); 
         return 0; 
      }; 
\end{verbatim}} 
 
Note how the information contained in the {\tt FLAG} has been used by the 
functions acting on the objects (in particular by the assignment operator). 
To optimize both speed and memory usage it is required for a C object (i.e. 
the copy made by the copy constructor of an F object existing at a higher 
level, see fig.~\ref{fig2} (right)) to be copied element by element into a 
new location of the memory, otherwise only its pointer is copied (the 
fastest way). Moreover when a new value is assigned to a C type object
it is promoted to type F. \\
As an exercise the reader is suggested to work out the deatils. \\
Once the class with the {\tt FLAG} has been implemented then all these 
technicalities can be forgotten but the safety rules, stated in the next
appedinx, should always be kept in mind!

\section{Safety rules}
 
Since the copy constructor has been redefined and it is now {\tt FLAG} 
dependent there are a few safety rules to follow to be sure that 
everything is working properly. These safety rules applys to objects
of class:
\begin{verbatim}
      Toy_class
      Matrix
      gauge_field 
      pl_field
      em_field
      fermi_field
\end{verbatim}

These safety rules are:
{\bf \begin{itemize} 
\item[$\bigstar$]
Always, before returning an {\it object}, 
change its {\tt FLAG} to R, i.e. call \\ {\tt prepare(}{\it object}{\tt);} 
 
\item[$\bigstar$]
Never explicitly use the copy constructor. 

\item[$\bigstar$]
Do not call the assignment operator of the argument of a function 
within the function itself (unless you understood completely how {\tt
Toy\_class} works).  
\end{itemize}} 
Attention: Different machines may have a different internal
representations of real numbers. Therefore some care must be taken when
copying data between different machines. This also holds
for the file containing the seed: \\ 
{\tt RandomBuffer.seed} 
\\ Therefore:
\begin{itemize}
\item[$\bigstar$] {\bf
Do not copy the seed from one machine to another, let the {\tt start();}
instruction create a new seed file.}
\end{itemize}
\newpage
\section*{Acknowledgemnts}

It is a pleasure to acknowledge C. Sachrajda, L. Del 
Debbio, J. Flynn, V. Lesk, J. Generowicz, K. Anderson (Southampton
University) for useful discussions and suggestions.
I also wish to thank C. Rebbi (Boston University) and C. Michael 
(Liverpool University) for letting me study their FORTRAN
montecarlo codes, from which I learned a lot.
The random number generator for real numbers encoded in {\tt
MDP\_Lib1.h} has been written by M. Beccaria and L. Del Debbio.


\begin{thebibliography}{99} 
\bibitem{core}  G. Satir and D. Brown, C++ The Core Language, O'Reilly and 
Associates (1995) 
 
\bibitem{bjarne}  B. Stroustup, C++ Programming Language, Addison-Welsey 
(1997) 

\bibitem{wilson} K. G. Wilson, Phys. rev. {\bf D10} (1974) 2445

\bibitem{symanzik} K. Symanzik, Nucl. Phys. {\bf B226} (1983) 187,205

\bibitem{sw} B. Sheikoleslami and R. Wolhert, Nucl. Phys. {\bf B259}
(1985) 572

\bibitem{creutz}  M. Creutz, Quarks, Gluons and Lattices, Cambridge
University Press (1983)

\bibitem{rothe}  Rothe, Lattice Gauge Theories, World Scientific 
 
\bibitem{montvay}  I. Montvay and G. M\"{u}nster, Quantum Fields on a 
Lattice, Cambridge 

\bibitem{me1}  M. Di Pierro and C. T. Sachrajda, A Lattice Study of Spectator
Effects in Inclusive Decays of B-Mesons, Nucl. Phys. {\bf B534} (1998);
hep-lat/9805028

\bibitem{me2} G. De Divitiis, L. Del Debbio, M. Di Pierro, J. Flynn,
C. Michael and J. Peisa, Towards a lattice determination of the
$B^\ast B \pi$ coupling, accepted for publication on JHEP (1998); 
hep-lat/9807032

\bibitem{me3} M. Di Pierro, Spectator Effects in Inclusive Decays of Beauty
Hadrons, to apper in the proceedings of Lattice98 (1998); hep-lat/9809083

\bibitem{me4} M. Di Pierro and C. T. Sachrajda, Spectator Effects in
Inclusive Decays of $\Lambda_b$ from Lattice Simulations (to be submitted).

\bibitem{cm}  N. Cabibbo and E. Marinari, A new method for updating $SU(N)$ 
matrices in computer simulations of gauge theories, Phys. Lett. {\bf
119B} (1982) 387 

\bibitem{errors} J. Shao and D. Tu, The Jackknife and Bootstrap,
Springer Verlag (1995)

\bibitem{lists}  M. Creutz, Quantum Fields on the Computer, World
Scientific (1992)

\bibitem{chris} C. T. Sachrajda, Lattice Simulations and Effective
Theories, Lectures presented at the Advanced School on Effective
Theories, Almunecar (Spain) June 1995; hep-lat/960527

\bibitem{heatlie} Heatlie et al., Nucl. Phys. {\bf B352} (1991) 266

\bibitem{stochastic1} G. Parisi, R. Petronzio and C. Rapuano,
Phys. Lett. {\bf B128} (1983) 418

\bibitem{michael} C. Michael and J. Peisa, Maximal variance reduction
     for stochastic propagators with applications to the static quark
     spectrum, Phys.Rev. {\bf D58} (1998) 

\bibitem{smearing} S. G\"usken {\it et al.}, Phys. Lett. B227 (1989) 266

\bibitem{metropolis}  G. Banhot, The Metropolis Algorithm, Rep. Prog. Phys. 
51 (1988) 429 

\bibitem{luscher} M. Luscher, Advanced Lattice QCD,
Talk given at Les Houches Summer School in Theoretical Physics,
Session 68, (1998); hep-lat/9802029
 
\bibitem{bernard} C. Bernard in ``From Action to Answers'',
Proceedings of the 1989 TASI in Elementary Particle Physics, Boulder,
World Scientific

\end{thebibliography}
\end{document}